\documentclass[
    paper=a4,
    pagesize=auto,
    fontsize=12pt,
    version=last,
    twoside=false,
    headings=small,
    ]
{scrartcl}
\pdfoutput=1

\usepackage[onehalfspacing]{setspace}
\KOMAoptions{DIV=14}

\usepackage[ngerman,english]{babel}
\usepackage[T1]{fontenc}
\usepackage[utf8]{inputenc}
\usepackage{lmodern}

\usepackage{float}

\usepackage{mathrsfs}
\usepackage{physics}

\usepackage[format = plain]{caption}
\usepackage{subcaption}

\usepackage{amsmath,amssymb,amsfonts}
\usepackage{graphicx}
\usepackage{xcolor}
\usepackage{diagbox}
\usepackage{abstract}
\usepackage[margin=1in]{geometry}

\usepackage[
    backend=biber,
    hyperref,
    doi,
    isbn=false,
    url=false,
    eprint=false,
    maxbibnames=6,
    giveninits=true,
    sorting=none,
    citestyle=numeric-comp]{biblatex}

\usepackage{csquotes}
\renewbibmacro{in:}{}

\AtEveryBibitem{
\clearfield{note}
\clearfield{pages}
\clearfield{month}
\clearfield{day}
\clearfield{location}
\clearfield{language}
\clearfield{number}}

\usepackage[
    colorlinks=true,
    citecolor={green!30!black},
    linkcolor={blue!50!black},
    urlcolor={blue!80!black}
]{hyperref}

\newcommand{\psp}{p_\text{sp}}

\begin{document}

\setlength{\footskip}{50pt}

\addtokomafont{disposition}{\sffamily}

\title{\textsf{Volumetric Benchmarking of Quantum Computing Noise Models}}

\date{\vspace{-7ex}}

\author{}

\maketitle

{\setlength{\parindent}{0pt}\large\sffamily Tom Weber$^{1}$, Kerstin Borras$^{2,3}$, Karl Jansen$^{4,5}$, Dirk Krücker$^{2}$, and Matthias Riebisch$^1(\dagger)$\par}
\vspace{3ex}
{\setlength{\parindent}{0pt}\sffamily
${}^{1}$Department of Informatics, University of Hamburg, Germany\\
${}^{2}$Deutsches Elektronen-Synchrotron DESY, Hamburg, Germany\\
${}^{3}$RWTH Aachen University, Aachen, Germany\\
${}^{4}$CQTA, Deutsches Elektronen-Synchrotron DESY, Zeuthen, Germany\\
${}^{5}$Computation-Based Science and Technology Research Center, The Cyprus Institute, Cyprus
}
\vspace{5ex}
\begin{abstract}
The main challenge of quantum computing on its way to scalability is the erroneous behaviour of current devices. Understanding and predicting their impact on computations is essential to counteract these errors with methods such as quantum error mitigation. Thus, it is necessary to construct and evaluate accurate noise models. However, the evaluation of noise models does not yet follow a systematic approach, making it nearly impossible to estimate the accuracy of a model for a given application. Therefore, we developed and present a systematic approach to benchmark noise models for quantum computing applications. It compares the results of hardware experiments to predictions of noise models for a representative set of quantum circuits.\par
We also construct a noise model and optimize its parameters with a series of training circuits. We then perform a volumetric benchmark comparing our model to other models from the literature.

\end{abstract}

\vspace{3ex}

\newpage
\thispagestyle{empty}
\vspace*{10cm}
\begin{center}
{\itshape\large To the memory of Matthias Riebisch.}
\end{center}

\newpage
\section{Introduction}
\label{sec:introduction}

Quantum computing is expected to offer novel applications in numerous fields of science. The most significant challenge to achieving scalable quantum computing is the level of errors in current noisy intermediate-scale quantum (NISQ) devices~\cite{Preskill2018}. Counteracting these errors is essential to enable reliable computations. While potential prospect solutions such as quantum error correction remain impracticable~\cite{Endo2018} due to small qubit numbers, quantum error mitigation methods can improve results significantly without causing an overhead of necessary qubit resources. Various methods aim to tackle different types of errors for several applications. Especially for algorithms like Variational Quantum Eigensolver (VQE)~\cite{Peruzzo2014}, we already find a large number of protocols trying to mitigate, e.g., readout error, gate error, or cross-talk~\cite{funcke_measurement_2022,Kwon2020a,Kandala2019,Temme2017, su_error_2021,Sun2021, Bravyi2021, Murali2020,vovrosh_simple_2021}. Almost all error mitigation methods have in common that they require additional quantum computing time~\cite{takagi_fundamental_2022}, which is limited. A prioritization of the dominant types of error is therefore necessary. Understanding and predicting the noisy behaviour of a quantum computer makes accurate noise models are indispensable for efficient and reliable quantum computing calculations.\par
However, there is no systematic approach for evaluating the quality of a noise model. Its accuracy is often estimated using a small number of arbitrary test circuits and comparing the model’s prediction to the results obtained with quantum hardware. These test circuits are usually not similar enough to realistic application circuits to allow for a generalization of the results. The size of the quantum circuits is usually too small, making it difficult to assess the accuracy of a noise model in a realistic application context.\par
Therefore, we propose a volumetric benchmarking approach to enable a systematic evaluation of quantum computing noise models. It is based on the framework presented in~\cite{blume-kohout_volumetric_2020}, which measures the performance of quantum computers. For a choice of representative quantum circuits, our benchmark compares the predictions of a noise model to results from quantum hardware. This procedure is carried out for different pairs $(w,d)$ of width $w$ and depth $d$ of the quantum circuits, where the width corresponds to the number of qubits, and the depth can be related to the number of consecutive gates or layers thereof. This allows evaluating the quality of a noise model as a function of the problem complexity, hence the name volumetric benchmark. The applications of our benchmarks are versatile. With a noise model assessed as being accurate, one can use simulations employing this model for test purposes or if quantum resources are limited. The knowledge gained through the model can also help prioritize quantum error mitigation methods.\par

In section~\ref{sec:methods}, we construct a noise model with trainable parameters that aims to represent the noise impact on VQE or similar algorithms. The parameters correspond to probabilities that certain errors occur and are optimized using the SPSA~\cite{Spall1992} algorithm. We then conduct volumetric benchmarks of the resulting noise model. The hardware experiments are run on IBM quantum hardware.

\subsection{Contribution}
\label{subsec:contribution}
This paper contains two main contributions:
\begin{enumerate}
    \item A benchmarking protocol for measuring the accuracy of quantum computing noise models, including a discussion regarding quality attributes from the systems benchmarking literature.
    \item The construction of a noise model, training of its parameters, and an evaluation with volumetric benchmarks using typical VQE quantum circuits on IBM quantum hardware. The model is compared to the \texttt{ibmq\_manila} device noise model provided in qiskit~\cite{Qiskit}.
\end{enumerate}

\subsection{Related work}
\label{subsec:relatedeork}
This section summarizes relevant research related to this paper. It discusses the literature on benchmarking for quantum hardware, the construction of noise models, and their calibration.\par

A variety of tomography approaches exists  for characterizing and benchmarking quantum computers~\cite{greenbaum_introduction_2015}. Quantum State Tomography (QST)~\cite{leibfried_experimental_1996} is a procedure that characterizes an unknown state $\rho$. This state could then be compared to the expected, ideal state of a quantum computation to obtain an estimate of the fidelity of the hardware. Quantum Process Tomography~ (QPT)\cite{chuang_prescription_1997,altepeter_ancilla-assisted_2003,dariano_quantum_2002, mohseni_quantum-process_2008, poyatos_complete_1997,shabani_efficient_2011} measures the process matrix of quantum gates. Both QST and QPT consider state preparation and measurements to work correctly. This is presently not always the case, making the estimates of quantum states and gates erroneous. In contrast, our noise model includes state preparation and measurement (SPAM) errors, and our training approach ensures they are represented appropriately. Gate Set Tomography (GST)~\cite{nielsen_gate_2021,merkel_self-consistent_2013, blume-kohout_robust_2013} also takes into account erroneous state preparation and measurements. While QPT characterizes a single gate, GST can reconstruct a set of operations in a self-consistent way. Many quantum experiments are needed to achieve this characterization, and scalability is problematic for benchmarking large systems. QPT and GST attempt to describe the noisy processes of a quantum device, but the process matrices do not give conceptual insights into the errors. This paper comprehensively constructs a noise model derived from the underlying physical processes.\par

Randomized benchmarking (RB)~\cite{emerson_symmetrized_2007,knill_randomized_2008,emerson_scalable_2005} measures the average gate error rates of a quantum computer. Many variations of RB exist, including cycle benchmarking~\cite{erhard_characterizing_2019}. These methods evaluate the performance of quantum hardware and do not attempt to describe the noise in detail. Moreover, they do not provide prospects on the impact of the noise.\par

In addition to all these tomography methods, there are other prominent benchmarking approaches for quantum hardware. Quantum volume~\cite{cross_validating_2019} is a single-number metric that indicates the maximal size of quantum circuits that can be executed successfully on a device. In~\cite{blume-kohout_volumetric_2020}, the authors propose a volumetric benchmarking approach that generalizes the quantum volume metric.  Volumetric benchmarks using mirror circuits are conducted in~\cite{proctor_measuring_2022} and applied to quantum error mitigation in~\cite{cirstoiu_volumetric_2022}. Our work transfers volumetric benchmarks to a different setting. While they originally measure the capabilities of quantum hardware for different problem sizes, we evaluate the accuracy of noise models in a volumetric manner and aim to provide prospects on the potential impact of that noise.\par

In~\cite{blume-kohout_wildcard_2020}, the authors present a wildcard error that accounts for inconsistencies between noise model predictions and hardware data. The amount of wildcard error needed can be interpreted as an estimate of the accuracy of the noise model. Machine learning methods are also used to describe quantum noise. In~\cite{Harper2020}, the authors propose a learning procedure to obtain the error rates of a quantum computer. In~\cite{georgopoulos_modelling_2021}, a noise model construction and its evaluation with test quantum circuits are presented. Our noise model adds crosstalk and a more advanced representation of readout error. Furthermore, we introduce a more systematic approach for benchmarking the noise models and present a different training approach for the model parameters.\par

A recent publication~\cite{gustiani_virtual_2023} presents a platform to emulate various types of quantum devices, such as superconducting qubits or trapped ions. Different noise models are used to simulate each of these devices. Some of these noise models overlap with the one we construct in this work. However, the specific models for readout and crosstalk error and the method to optimize the noise model parameters differ. Moreover, there is no systematic evaluation of the presented noise models.

\section{Background}
\label{sec:background}

This section introduces the different types of noise that can occur on a quantum computer, our notation, and quality criteria for benchmarks. 

\subsection{Quantum Noise and Noise Models}
\label{subsec:quantumnoise}

Quantum noise refers to all interactions of a quantum system with its environment. In quantum computing, these interactions lead to erroneous computations. Not only do qubits interact with their environment, but also with each other. Therefore, quantum circuits are not executed as intended. For instance, measurements can be faulty (readout error), or gates are applied imperfectly (gate error).\par

Noise models offer means of describing and predicting the noisy effects in a quantum device. They contain information about the error types and the point in a quantum circuit where they occur. More precisely, a noise model maps quantum circuits to outcome probability distributions~\cite{blume-kohout_wildcard_2020}. One would obtain these distributions by running the circuits many times on a noisy quantum computer that behaves the way the model describes.\par
Quantum operations~\cite{nielsen_quantum_2010} on density matrices (or density operators) are the prevalent mathematical formalism to model quantum noise. This section does not give detailed mathematical definitions of quantum operations but focuses on specific examples of noise channels. We refer to appendices~\ref{sef:densityoperators} and~\ref{sec:quantumoperations} instead for more mathematical details. A comprehensive discussion of the subject can also be found in~\cite{nielsen_quantum_2010}.\par

In the following, let $\rho$ be a density matrix describing the state of a set of qubits and denote the Pauli matrices as
\begin{align*}
    \mathsf{X} = \begin{pmatrix}
    0 & 1\\
    1 & 0
    \end{pmatrix},\quad
    \mathsf{Y} = \begin{pmatrix}
    0 & -i\\
    i & 0
    \end{pmatrix},\quad
    \mathsf{Z} = \begin{pmatrix} 
    1 & 0\\
    0 & -1
    \end{pmatrix}\ .
\end{align*}
\subsubsection*{Readout error}

\begin{figure}
\centering
\includegraphics[width = 0.5\textwidth]{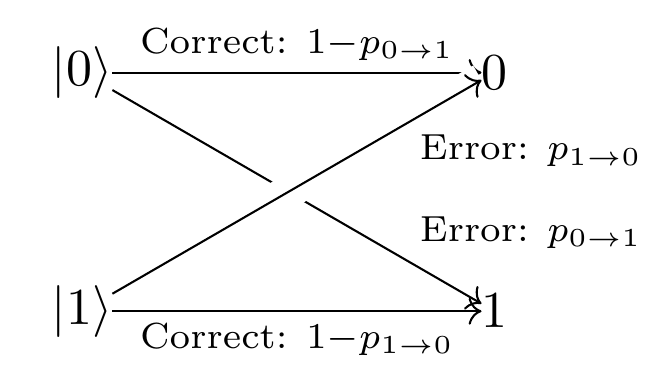}
\caption{Graphical representation of readout error on a single qubit with different bit-flip probabilities $p_{0\to1}$ and $p_{1\to0}$.}
  \label{fig:readouterror}
\end{figure}

The measurement of qubits on current quantum computers is often erroneous, with error rates of up to 30\%~\cite{Tannu2019a}, although this has improved on the newest machines. This behaviour is called measurement error or readout error and can be modelled as a classical bit-flip as follows~\cite{funcke_measurement_2022}: For each qubit $q$, a measurement outcome '0' is mistakenly recorded as '1' with probability $p_{0\to1}(q)$ and vice versa with $p_{1\to0}(q)$ (as shown in figure~\ref{fig:readouterror}). Note that the probabilities can be asymmetric. In this work, we use the terms readout error and measurement error interchangeably.\par

Several papers have been published on readout errors and methods to mitigate them, particularly for measuring expectation values of observables on a quantum computer. Obtaining these expectation values is an essential step in VQE algorithms~\cite{Peruzzo2014}.

\subsubsection*{State preparation error}

At the beginning of a quantum circuit, the qubits of a quantum computer are prepared in an initial state. Typically, the state $\rho_0=\dyad{0\cdots 0}$ is chosen, where $\ket{0\cdots 0}=\ket{0}^{\otimes N}$ and $N$ denotes the number of qubits. This procedure can be imperfect, resulting in an incorrect initial state and, thereby, unreliable computation with low fidelity.\

When the initial state is given as above, the state preparation error can be modelled as applying an  $\mathsf{X}$ gate on qubit $q$ with probability $\psp(q)$~\cite{geller_toward_2021}. For a single qubit, this yields the following noise channel:

\begin{align}
    \rho_0\mapsto (1-\psp(q))\rho_0 + \psp(q)\cdot \mathsf{X}\rho_0\mathsf{X}\ .
    \label{eq:statepreparation}
\end{align}

\subsubsection*{Depolarizing error}

Besides state preparation and measurement, gate operations are imperfect. On current devices, both one and two-qubit gates are affected, where the error rates of two-qubit gates are usually higher~\cite{sanders_bounding_2015}.\par
Depolarizing error is an important type of gate error. A qubit is depolarized if its state is completely mixed and all information is lost. In terms of density matrices, the state $\rho$ is replaced by the normalized identity matrix $\mathsf{I}/D$ with probability $\lambda$, where $D$ is the dimension of the quantum system, i.e., $D=2^N$ for depolarization of $N$ qubits. The depolarizing channel can be written as
\begin{align}
\label{eq:depolarization}
    \mathcal{D}(\rho) = (1-\lambda)\rho+\frac{\lambda}{D}\cdot\mathsf{I}\ .
\end{align}
Following~\cite{nielsen_quantum_2010}, one can rewrite the equation above for a single qubit as
\begin{align*}
    \mathcal{D}(\rho) = (1-\frac{3\lambda}{4})\rho + \frac{\lambda}{4}(
    \mathsf{X}\rho\mathsf{X}
    +\mathsf{Y}\rho\mathsf{Y}
    +\mathsf{Z}\rho\mathsf{Z})\ .
\end{align*}
If depolarization affects a gate $g$ on a single qubit $q$ (or a pair of qubits $q_1,q_2$), we denote the probability by $\lambda_g(q)$ (or $\lambda_g(q_1,q_2)$).

\subsubsection*{Thermal relaxation and dephasing}
As a qubit interacts with its environment, it is subject to two central dynamics: thermal relaxation towards its ground state and dephasing~\cite{georgopoulos_modelling_2021, zurek_decoherence_1991}. Assuming that the qubit is realized with $\ket{0}$ as its energetic ground state, thermal relaxation refers to the decay towards $\ket{0}$ over time. The mean lifetime of that decay is commonly labelled $T_1$.\par

Moreover, a qubit experiences a decay towards classical behaviour called dephasing. Similarly to thermal relaxation, this decay is determined by the time $T_2$. The times $T_1$ and $T_2$ are related by $T_2\leq 2\cdot T_1$.\par

For simplicity, we first assume $T_2<T_1$. In that case, thermal relaxation can be modelled as a reset operator $\dyad{0}$ that acts on the density matrix $\rho$ with probability $p_{\text{reset}}$~\cite{georgopoulos_modelling_2021}. During quantum computation, the probability depends on the time $T_g$ it takes to apply a gate operation $g$ to the qubits. It is given by $p_{\text{reset}}=1-\exp(-T_g/T_1).$\par

Dephasing can be modelled as the Pauli $\mathsf{Z}$ operator acting with probability $p_\mathsf{Z}$. This probability is computed from the times $T_1$, $T_2$, and $T_g$ by~\cite{georgopoulos_modelling_2021}
\begin{align*}
    p_\mathsf{Z} = \frac{(1-p_{\text{reset}})(1-\exp(-T_g/T_2+T_g/T_1))}{2}\ .
\end{align*}
Thus, the noise channel representing thermal relaxation and dephasing can be written as
\begin{align}
    \mathcal{T}(\rho) = p_\mathsf{I}\rho + p_\mathsf{Z}\cdot  \mathsf{Z}\rho\mathsf{Z} +
    p_{\text{reset}}\cdot \dyad{0}\rho \dyad{0},
    \label{eq:thermal_relaxation}
\end{align}
where $p_\mathsf{I} = 1-p_\mathsf{Z}-p_{\text{reset}}$. If $T_2>T_1$, one cannot write thermal relaxation and dephasing as above but must switch to a representation by a Choi matrix~\cite{choi_completely_1975} instead. A detailed discussion can be found in~\cite{georgopoulos_modelling_2021}.\par

In our situation, we write $T_{1,2}(q)$ for the $T_{1,2}$ time corresponding to qubit $q$. For a two-qubit operation on qubits $q_1$ and $q_2$, thermal relaxation and dephasing are considered to be two instances of single-qubit thermal relaxation and dephasing with the respective parameters $T_{1,2}(q_1)$ and $T_{1,2}(q_2)$.

\subsubsection*{Crosstalk error}
The error types discussed above consider the interactions of a qubit with its environment to be local and independent of other qubits. In reality, many processes violate locality or independence. These processes are called crosstalk and lead to crosstalk errors. In this paper, we only consider a basic model of crosstalk error. An extensive discussion can be found in~\cite{sarovar_detecting_2020}.\par

We represent crosstalk error as follows. Each time an erroneous single-qubit gate $g\in\lbrace\mathsf{X},\sqrt{\mathsf{X}}\rbrace$ is applied to qubit $q$, it causes a rotation
\begin{align}
\label{eq:crosstalk}
    \mathsf{R}_x(\phi) = \exp\left(-i\frac{\phi_g(q)}{2}\mathsf{X}\right)
\end{align}
on its neighbour qubits, where $\phi_g(q)$ is the rotation parameter.

\subsection{Benchmarking Quality Criteria}
\label{subsec:qualitycriteria}
Comprehensive literature exists on benchmarking classical computing systems or components. In the following, we review the most important aspects relevant to this paper. In~\cite{v_kistowski_how_2015}, a benchmark is defined as a tool for evaluating or comparing systems according to specific characteristics. These characteristics can be assigned to one of these categories of \textit{quality criteria}:
\begin{itemize}
    \item \textbf{Relevance}: A benchmark is relevant if it measures the behaviour of a system well, and its results can be generalized to real-world scenarios. In our situation, a benchmark should allow an estimation of the accuracy of a given noise model for applications of interest.

    \item \textbf{Reproducibility}: The results of a reproducible benchmark are consistent over multiple runs with the same configuration.

    \item \textbf{Fairness}: A benchmark is fair if it does not impose artificial constraints on the system under test. Hence, a benchmark for noise models should not favour one model over another a priori.

    \item \textbf{Verifiability}: The verifiability of a benchmark ensures it is performed correctly and instructions are respected. One measure of improving verifiability is self-validation.
    \item \textbf{Usability}: A benchmark is usable if it is easy to run by a user. The necessary hardware and software configuration for the benchmark should be straightforward to obtain.
\end{itemize}

\section{Volumetric Benchmarks for Noise Models}
In this section, we present our volumetric benchmarking approach for evaluating the accuracy of noise models. First, we explain this process in detail and discuss the differences to the framework explained in~\cite{blume-kohout_volumetric_2020}. Afterwards, we describe how improvements in the quality of such benchmarks in terms of the quality attributes from section~\ref{subsec:qualitycriteria} can be achieved.

\subsection{The Framework}
\label{subsec:framework}
A volumetric benchmark in our approach is always related to a noise model and a quantum device. For a collection of test circuits, it compares the model predictions to the results of the quantum device. A volumetric benchmark consists of the following steps:
\begin{enumerate}
    \item \textbf{Test circuits:} For pairs of width $w$ and depth $d$, define a set $C(w,d)$ of quantum circuits. These circuits are used to compare the results predicted by the noise model to hardware results from the device. The depth could correspond to the number of gates or the number of layers thereof, while the width $w$ is the number of qubits.
    \item \textbf{Compilation rules:} Set up rules for compiling the quantum circuits from step 1 to the native gates of the device, enabling it to run the circuits later. There are different methods to compile quantum circuits. Sometimes, it is more feasible to optimize the circuits during compilation. In other cases, one might be more interested in rules restricting this optimization.
    \item \textbf{Model predictions:} Specify a way to obtain the noise model’s predictions for the compiled quantum circuits. Among other things, such predictions could be made from noisy simulations of the circuits or exact computations using density matrices. We want to emphasize that the noise model must predict the results for the compiled circuits to allow for a meaningful comparison to hardware results. Further details are discussed later.
    \item \textbf{Hardware results:} Run the compiled quantum circuits on the quantum computer. The exact specifications of this run, e.g., order of execution or number of shots, need to be described in detail for better reproducibility.
    \item \textbf{Single circuit evaluation:} Define a metric that measures the difference between model prediction and hardware results for a single quantum circuit. For example, this metric could directly compare the outcome distributions of the circuit, or it could be based on higher-level attributes like expectation values of quantum mechanical observables.
    \item \textbf{Overall evaluation:} If the set $C(w,d)$ contains more than a single quantum circuit, specify how to derive an overall evaluation. For example, when a single circuit is assessed using the difference in observable expectations, this overall evaluation could be chosen as the average difference.
\end{enumerate}

While the approach presented in \cite{blume-kohout_volumetric_2020} compares hardware results to the ideal outcomes of quantum circuits, our approach uses hardware results and noise model predictions. Therefore, the third step is a novel extension of the original volumetric benchmarks. Moreover, our goal here is entirely different as we do not evaluate the behaviour of quantum devices but the accuracy of noise models.

\subsection{Ensuring quality}
\label{subsec:ensuringquality}
In section \ref{subsec:qualitycriteria}, we discuss benchmarking quality criteria such as relevance or reproducibility. The quality of our volumetric benchmarking framework for noise models in terms of these criteria depends on user choices during the different steps of a benchmark. This section discusses possibilities to increase the quality by examining the criteria individually. We first highlight potential obstacles for each attribute and explain how to avoid or minimize these.

\subsubsection*{Relevance}
Often, relevance is the essential quality attribute of a benchmark. Even if it perfectly meets all other attributes perfectly, it can be useless because its results are not transferrable to any real-life application of interest. For example, a noise model might perform well in a benchmark but be inaccurate at predicting the noisy hardware behaviour when used for an application such as VQE.\par
The relevance of our volumetric benchmarks strongly depends on the quantum circuits and evaluation criteria used. Typical quantum circuits should be chosen if one is interested in noise models for a particular field of application. Moreover, the more quantum circuits are used and the more they differ from each other, the more transferrable the benchmark results are. For parametrizable quantum circuits, this means that several sets of parameters should be used to avoid a dependency on a particular choice. The scalability can be increased by testing many configurations $(w,d)$ of width and depth.

\subsubsection*{Reproducibility}
Quantum computations and, thus, also volumetric benchmarks are naturally subject to fluctuations due to finite shot numbers. The consequence is that reproducibility can suffer because running the same test circuits can yield different results. Since quantum hardware also often exhibits a time drift, its performance depends on when it is used. Hardware calibrations can have a significant impact on the noise of a device.\par
All benchmark experiments should be run in as small a time window as possible to mitigate these effects. Moreover, the circuits should be run as often as possible to decrease the variance of their outcome. If the model predictions are obtained by simulations, they should also be performed with large shot numbers. Alternatively, one could use exact predictions based on density matrix computations to minimize fluctuations further.

\subsubsection*{Fairness}
Since our approach aims to benchmark noise models, artificial constraints on their performance are unlikely. If two noise models are supposed to predict the hardware behaviour for entirely different application contexts, they should not be compared in the first place. Fairness is mainly threatened when models are benchmarked at different times, and the hardware shows different levels of noise such that converging predictions can be more challenging. Again, this can be mitigated by performing benchmarks in a short time window and more often.

\subsubsection*{Verifiability}
The verifiability of a benchmark measures to what extent it runs as expected (see~\cite{v_kistowski_how_2015}). Optimally, a verifiable benchmark includes some self-validation. For quantum circuits run on noisy hardware, such self-validation is challenging to achieve.\par

One could run simple circuits of which the ideal results are known and ensure that the hardware results are within a reasonable deviation range. The model predictions could also be validated by comparing exact computations to noisy simulations.

\subsubsection*{Usability}
The main threat to usability is restricted access to quantum hardware, so not everyone can easily perform a volumetric benchmark. Publicly providing quantum experiments data can help researchers run benchmarks with that data to test their noise models.

\section{Methods}
\label{sec:methods}
This section describes the construction, training, and evaluation of a noise model for quantum computing. The noise model incorporates erroneous state preparation and measurement, depolarization, thermal relaxation, and crosstalk error. It is inspired by the model provided in~\cite{georgopoulos_modelling_2021} and depends on a set of parameters determined by training the model in a machine learning-like fashion. A detailed introduction of the noise model and the parameter optimization is given later.\par

This section is structured as follows. Firstly, we construct the noise model without specifying its parameters. It is defined on the native gate set of the IBM Quantum Falcon processors. Secondly, we explain the training procedure, including the definition of training and test sets, as well as the choice of an optimization algorithm. Finally, we benchmark the resulting and other noise models, particularly the device noise model provided in qiskit.

\begin{figure}
    \centering
    \includegraphics[width = 0.6\textwidth]{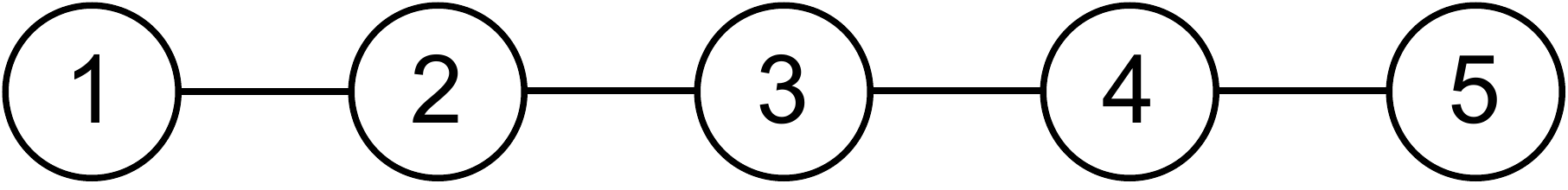}
    \caption{Qubit layout for IBM's \texttt{ibmq\_manila} device.}
    \label{fig:qubit_layout}
\end{figure}

\subsection{Constructing the noise model}
\label{subsec:construction}

The noise model that we use later for training and benchmarking describes quantum computers similar to IBM's \texttt{ibmq\_manila} device using a Falcon processor. It can be easily generalized to any other gate-based quantum device with few minor adaptions. The \texttt{ibmq\_manila} machine has $N=5$ qubits in a linear layout (see figure~\ref{fig:qubit_layout}). It implements three single-qubit gates ($\mathsf{X}$,$\mathsf{\sqrt{X}}$,$\mathsf{R}_z$) and one two-qubit gate ($\mathsf{CNOT}$) as native gates. We denote the native gate set by $\mathcal{G}=\lbrace\mathsf{X}, \mathsf{\sqrt{X}},\mathsf{R}_z,\mathsf{CNOT}\rbrace.$

\begin{figure*}
    \centering
    \includegraphics[width=\textwidth]{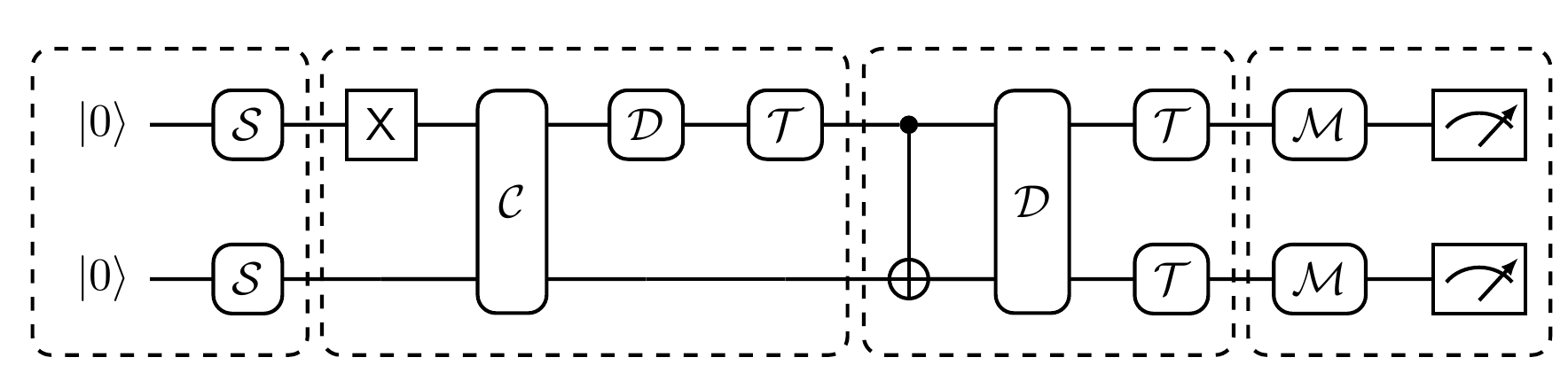}
    \caption{Example of a quantum circuit subject to our noise model. $\mathcal{S}$, $\mathcal{C}$, $\mathcal{D}$, $\mathcal{T}$, and $\mathcal{M}$ denote state preparation, crosstalk, thermal relaxation, depolarization, and measurement error, respectively. Note that the crosstalk error is depicted on both qubits to emphasize their interaction. The dashed boxes indicate what circuit operations are affected by which errors.}
    \label{fig:noisy_circuit}
\end{figure*}

Our noise model can describe any device with $N$ qubits in a linear layout. For other layouts, adaptions must be made to the possible multi-qubit interactions. The model combines all types of noise defined in section~\ref{subsec:quantumnoise}. At the beginning of each computation, the initial state is prepared as $\rho_0=\dyad{0\cdots0}$. It is followed by state preparation error $\mathcal{S}$ with corresponding probabilities $\psp(q)$, yielding $N$ model parameters.\par

Afterwards, gates are applied to the resulting (possibly erroneous) state. Each gate $g$ is followed by
\begin{itemize} 
    \item crosstalk error $\mathcal{C}$ (for $g\in\lbrace\mathsf{X},\sqrt{\mathsf{X}}\rbrace$) with parameter $\phi_g(q)$, applied to the neighbour qubits,
    \item depolarizing error $\mathcal{D}$ with parameters $\lambda_g(q)$ for single-qubit gates and $\lambda_g(q_1,q_2)$ for the $\mathsf{CNOT}$ gate, applied to the gate qubit(s),
    \item and thermal relaxation and dephasing $\mathcal{T}$ with parameters $T_{1,2}(q)$, applied to the gate qubit(s).
\end{itemize}

After all gates and their errors have been applied, measurement error $\mathcal{M}$ affects all qubits with parameters $p_{0,1\to1,0}(q)$. Finally, the qubits are measured in the computational basis.
A basic example of our noise model on a quantum circuit containing only one $\mathsf{X}$ gate and one $\mathsf{CNOT}$ gate can be found in figure~\ref{fig:noisy_circuit}.\par

The total number of parameters of our noise model for a system of $N$ is $11N-1$, as shown in table~\ref{tab:numberofparameters}. Note that we assume the $\mathsf{CNOT}$ gates to only be applied in one direction per qubit pair. Since there are three types on one-qubit gates and one two-qubit gate, there are $4N-1$ model parameters corresponding to depolarization error.

\subsection{Simulating the noise model}
\label{subsec:simulatingthenoisemodel}
All noisy simulations using the above model are carried out by exactly computing the density matrix of the system and its change due to errors. As section~\ref{subsec:quantumnoise} explains, all errors included in the model can be represented by quantum operations, which are linear maps of the density matrix. These linear maps depend on the respective parameters of the errors, e.g., bit-flip probabilities in the case of readout error. Pennylane~\cite{https://doi.org/10.48550/arxiv.1811.04968} offers the possibility to implement the errors and simulate quantum circuits with our noise model.\par

\begin{table*}
    \centering
    \begin{tabular}{|c|c|c|c|}
    \hline
    symbol & error & parameters & number of parameters\\
    \hline
    $\mathcal{S}$  & state preparation & $\psp(q)$ & $N$\\
    \hline
    $\mathcal{D}$  & depolarization & $\lambda_g(q)$ & $4N-1$\\
    \hline
    $\mathcal{C}$  & crosstalk & $\phi_g(q)$ & $2N$\\
    \hline
    $\mathcal{T}$  & thermal relaxation & $T_{1,2}(q)$ & $2N$\\
    \hline
    $\mathcal{M}$  & measurement & \begin{tabular}{@{}c@{}}$p_{0\to1}(q)$, \\ $p_{1\to0}(q)$\end{tabular} & $2N$\\
    \hline
    \hline
    total & & & $11N-1$\\
    \hline
    \end{tabular}
    \caption{Number of parameters corresponding to each type of error in the noise model, assuming a system of $N$ qubits. The parameters are later optimized during model training. See figure \ref{fig:readouterror} and (\ref{eq:statepreparation})-(\ref{eq:crosstalk}) for more details.}
    \label{tab:numberofparameters}
\end{table*}

After initializing the density matrix, all gates and errors are applied, and the final density matrix is computed. From this final density matrix, we obtain one of the following two quantities: in training, we compute the outcome distribution of basis states, while for the benchmarks, we use the expectation value of the $\mathsf{Z}^{\otimes w}$ operator.\par

If we take the situation from figure~\ref{fig:noisy_circuit} as an example, the initial density matrix is $\rho_0=\dyad{00}$. Afterwards, (\ref{eq:statepreparation}) is used to apply state preparation error with probability $\psp(0)$ to the first and with probability $\psp(1)$ to the second qubit. Next, the $\mathsf
X$ gate acts on the first qubit, followed by crosstalk, depolarizing, and thermal relaxation error. These errors are computed using the corresponding equations from section~\ref{subsec:quantumnoise} with parameters $\phi_\mathsf{X}(0)$, $\lambda_\mathsf{X}(0)$, and $T_{1,2}(0)$, respectively. The rest of the computation is done similarly.

\subsection{Training the noise model}
\label{subsec:training}
This section explains how the parameters of our noise model from section~\ref{subsec:construction} can be optimized. The approach is inspired by machine learning in the sense that model predictions are repeatedly evaluated on a training data set, and parameters are adapted accordingly.\par

For the training, quantum circuits are first compiled to the native gate set $\mathcal{G}$. The compiled circuits are then both run on a quantum computer and simulated with the noise model. Afterwards, a loss function is defined that measures the deviation of model predictions from hardware outcomes of the quantum circuits. Details on the implementation can be found in appendix~\ref{sec:implementation}.

\subsubsection*{Training set}
The training set contains 100 quantum circuits for which the noise model predicts noisy outcomes. Since we are mainly interested in VQE quantum computing applications, these training circuits consist of alternating layers of single-qubit rotations and entanglement. They follow the \texttt{EfficientSU2} structure that is part of the qiskit library. The number of layers is denoted as $d$.\par

\begin{figure*}[t]
    \centering
    \includegraphics[width=0.8\textwidth]{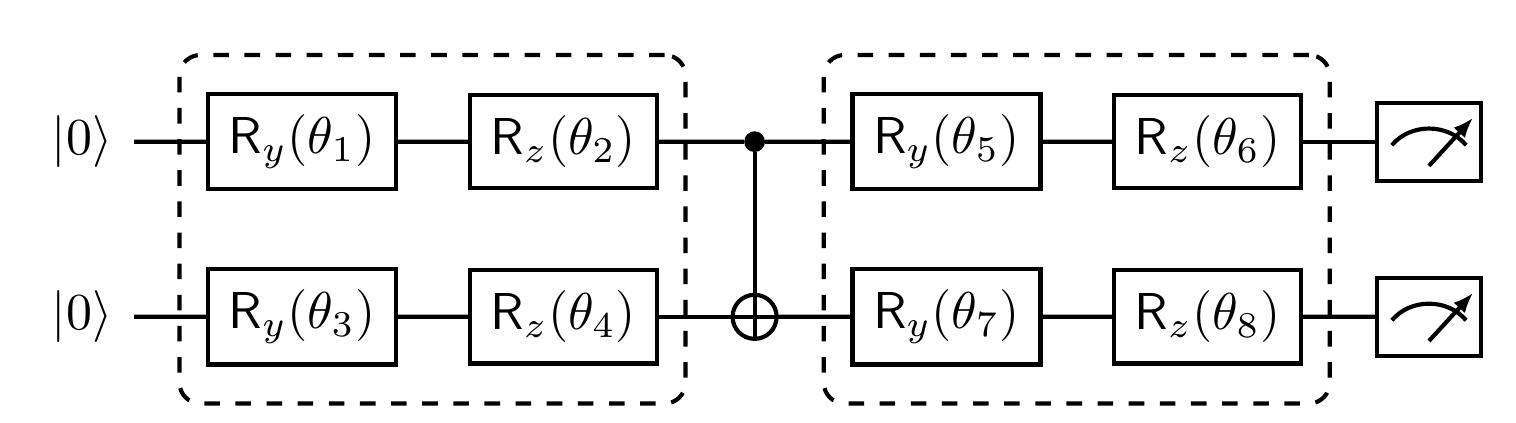}
    \caption{\texttt{EfficientSU2} circuit with three layers on two qubits. Each rotational layer consists of an $\mathsf{R}_y$ gate followed by an $\mathsf{R}_z$ gate on each qubit. Entanglement layers implement $\mathsf{CNOT}$ gates acting linearly on all qubit pairs.}
    \label{fig:efficientsu2}
\end{figure*}

Each rotational layer applies an $\mathsf{R}_y$ gate followed by an $\mathsf{R}_z$ gate to each qubit. Their rotation angles are randomized. The entanglement layers consist of $\mathsf{CNOT}$ gates that linearly connect all qubits. For example, the resulting quantum circuit for $d=3$ layers and $w=2$ qubits is shown in figure~\ref{fig:efficientsu2}.

\subsubsection*{Loss function}
The loss function compares the model predictions to the outcomes of the quantum circuits run on the device. The latter ones are measurement counts of computational basis states. Exact density matrix simulations allow to compute the probability distribution of these measurements based on the noise model. If one interprets the hardware counts as relative frequencies, the task is to compare two probability distributions.\par

Various metrics measure the distance between two distributions, e.g., Kullback-Leibler (KL) divergence~\cite{kullback_information_1951}. Here, we use the Hellinger distance~\cite{hellinger_neue_1909} between the probability distributions from the simulation and hardware run to define the loss on a single quantum circuit. If $P=(p_i)_{i\in\mathcal{I}}$ and $Q=(q_i)_{i\in\mathcal{I}}$ are two discrete probability distributions, then their Hellinger distance $H(P,Q)$ is defined as
\begin{align}
    H(P,Q) = \frac{1}{\sqrt{2}}\sqrt{\sum\limits_{i\in\mathcal{I}} (\sqrt{p_i}-\sqrt{q_i})^2}\ .
\end{align}
For a set of multiple circuits, as in training, we define the loss function as the arithmetic mean of Hellinger distances. Training the noise model with Kullback–Leibler divergence leads to similar results.

\subsubsection*{Optimization algorithm}
For optimizing the parameters of our noise model, we use the simultaneous perturbation stochastic approximation (SPSA) algorithm~\cite{Spall1992}. Since each evaluation of the loss function involves simulating 100 quantum circuits, the optimizer must should only a few evaluations to be cost-effective. The SPSA algorithm approximates gradients with only two evaluations of the loss function per iteration. Therefore, it is well suited for this task.

\subsection{Other noise models}
\label{subsec:othernoisemodels}

Besides the trained noise model from above, we benchmark two others. The first one only includes readout error, where the bit-flip probabilites are obtained from the device calibration by IBM. It serves as a basic example to explain our approach here and can be simulated exactly using density matrices.\par

The second other noise model is the \texttt{ibmq\_manila} device noise model provided in qiskit. Since its calibration changes with time, we always use the corresponding snapshots of the model when comparing it to hardware data. Moreover, simulations with the device noise model are always done shot-wise, meaning that we cannot compute predictions exactly. We mitigate possible variance effects by using large shot numbers.\par

In summary, we consider the following three noise models:

\begin{itemize}
    \item \textbf{Readout model}: only includes readout error with bit-flip probabilites from the IBM calibration.
    \item \textbf{Device model}: qiskit noise model for the \texttt{ibmq\_manila} quantum computer. Consists of readout error, depolarization, and thermal relaxation.
    \item \textbf{Trained model}: includes readout error, depolarization, thermal relaxation, cross error, and state preparation error. Its parameters are not determined by hardware calibration, but by optimization using training circuits.
\end{itemize}

\subsection{Volumetric benchmarks}
\label{subsec:volumetricbenchmarks}
This section describes how the volumetric benchmarks of different noise models were conducted. It follows the procedure from section~\ref{subsec:framework}. For the benchmarks, the \texttt{ibmq\_manila} device was used. Therefore, the native gate set is $\mathcal{G}=\lbrace\mathsf{X}, \mathsf{\sqrt{X}},\mathsf{R}_z,\mathsf{CNOT}\rbrace$, and the maximal number of qubits for the benchmark is $w=5$. Further implementation details can be found in appendix~\ref{sec:implementation}.\par

\begin{enumerate}
    \item \textbf{Test circuits.} Similarly to the training, we use \texttt{EfficientSU2} circuits for benchmarking. They have alternating layers of rotational and entanglement gates. All odd layers consist of an $\mathsf{R}_y$ gate followed by an $\mathsf{R}_z$ gate, starting with the first layer. In between, there are $\mathsf{CNOT}$ gates that linearly connect all qubits.\par
    
    For width $w$ and depth $d$, the set $C(w,d)$ consists of 200 circuits with $d$ layers acting on $w$ qubits. The rotation angles are randomized for each circuit. Note that we use different circuits for training and benchmarking.
    
    \item \textbf{Compilation rules.} Different types of optimization can be applied during the compilation of quantum circuits. With no optimization, every gate of the circuit is compiled into a representation by native gates. Otherwise, the number of gates in the resulting circuit is minimized to reduce the impact of quantum noise.\par
    The qiskit library offers several configurations for this process, which are applied by the \texttt{transpile} function and its \texttt{optimization\_level} argument. We choose \texttt{optimization\_level = 2} for compilation. 
    
    \item \textbf{Model predictions.} Exact simulations based on density matrices are used here to predict noisy $\mathsf{Z}^{\otimes w}$ expectations values of the test circuits, see section~\ref{subsec:simulatingthenoisemodel} and appendix~\ref{sec:implementation} for further information.\par
    If one wants to benchmark the device noise model from qiskit, only shot-wise simulations of the circuits are supported. Therefore, the outcomes are measurement counts of basis states, as in the case of hardware results. We use 8192 shots for each simulation, which is sufficient to reduce shot noise, i.e., statistical uncertainty due to finite shot numbers.
    
    \item \textbf{Hardware results.} The hardware results of the test circuits are obtained with the \texttt{ibmq\_manila} device. As for the simulations above, every circuit is run 8192 times to reduce shot noise.\par
    
    Since the jobs are placed in a queue, the experiments for different pairs $(w,d)$ cannot be run simultaneously. We save a snapshot of the device noise model before each run to enable its fair evaluation later. Hence, comparing different noise models does not depend on the execution time. The running times for each experiment can be found in table~\ref{tab:running_times}.
    
    \item \textbf{Single circuit evaluation.} To compare the model predictions to the hardware results for a single quantum circuit $c\in C(w,d)$, we evaluate the expectation values of the $\mathsf{Z}^{\otimes w}$ operator and compute their absolute difference
    \begin{align}
        d(c)=\abs{\expval{\mathsf{Z}_{\text{model}}^{\otimes w}(c)}-\expval{\mathsf{Z}_{\text{hardware}}^{\otimes w}(c)}}\ .
        \label{eq:singleloss}
    \end{align}
    
    \item \textbf{Overall evaluation.} We compute the arithmetic mean of all single-circuit results for the overall evaluation: 
    \begin{align*}
        L = \frac{1}{n}\sum\limits_{c\in C(w,d)}d(c)\ ,
    \end{align*}
    where $n$ is the number of circuits in $C(w,d)$.
\end{enumerate}

\subsection{Confidence intervals}
\label{subsec:confidenceintervals}

To estimate the statistical significance of our results, we perform bootstrapping and compute confidence intervals based on the resulting bootstrap distributions. The bootstrapping procedure is as follows. Firstly, we randomly sample 200 quantum circuits with replacement from the set of test circuits $C(w,d)$. The hardware (and qiskit simulation) results of a single quantum circuit consist of 8192 shots. Secondly, we sample 8192 shots with replacement for each circuit. This sampling process is repeated 1000 times to obtain the bootstrap distribution. Finally, we compute the 95\% confidence intervals based on the percentiles of this distribution.

\section{Results}
\label{sec:results}

\begin{figure}
    \centering
    \includegraphics[width=0.5\textwidth]{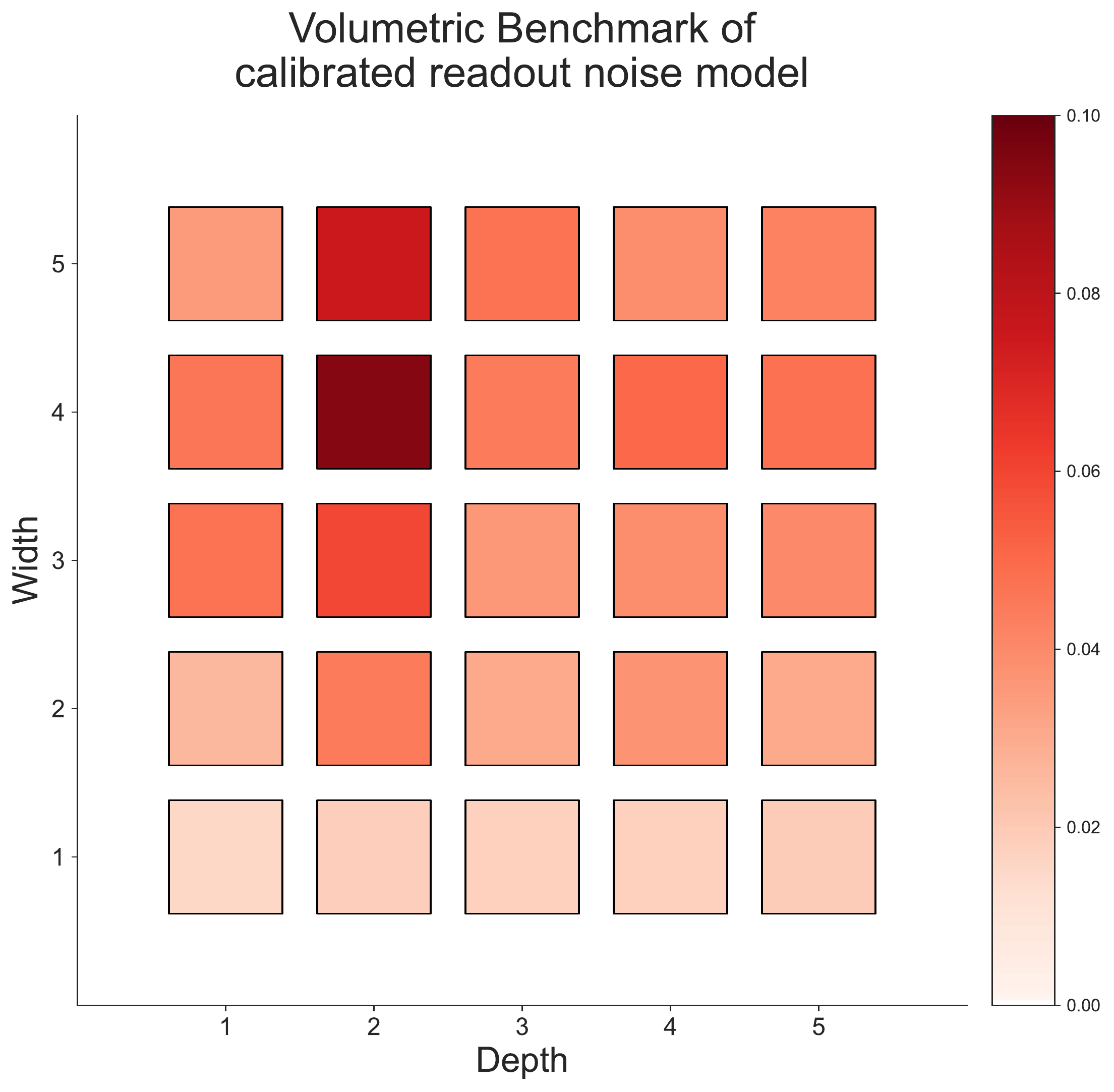}
    \caption{Volumetric benchmark results for readout noise model with calibrated bit-flip probabilities from section \ref{subsec:othernoisemodels}. The colours represent the average absolute deviation of the predicted $\mathsf{Z}^{\otimes w}$ expectation value from the hardware data. Darker squares indicate larger deviations and worse model accuracy.}
    \label{fig:result_readout}
\end{figure}

This section presents the results of volumetric benchmarks for the noise models from above. Details on the benchmark process and noise models are given in section~\ref{sec:methods}. Recall that the volumetric benchmark compares $\mathsf{Z}^{\otimes w}$ expectation values of noisy simulation and hardware experiment for different widths $w$ and depths $d$. The result of each configuration $(w,d)$ is represented by a different square in the figure. The overall style of presentation is inspired by~\cite{blume-kohout_volumetric_2020}.\par

The colour of a square indicates the average absolute deviation between noise model prediction and hardware data. Darker squares indicate a larger deviation, while white squares indicate good agreement. On the right side of the plot, one can find a legend explaining how the colours translate to numeric values. This legend is valid for all three plots, so the benchmark results for all noise models can be directly compared.\par

Consider the readout model and its benchmark results in figure~\ref{fig:result_readout} as an instructive example. The figure shows that the noise model predicts the hardware behaviour well for $w=1$, i.e., for a single qubit. For larger qubit numbers, the deviations between model predictions and hardware data increase quickly. For example, one finds an average absolute error of $\mathsf
{Z}^{\otimes w}$ of almost 0.1 for $w=4,d=2$.\par

\begin{figure*}
\centering
     \begin{subfigure}[b]{0.45\textwidth}
         \centering
         \includegraphics[height=0.3\textheight]{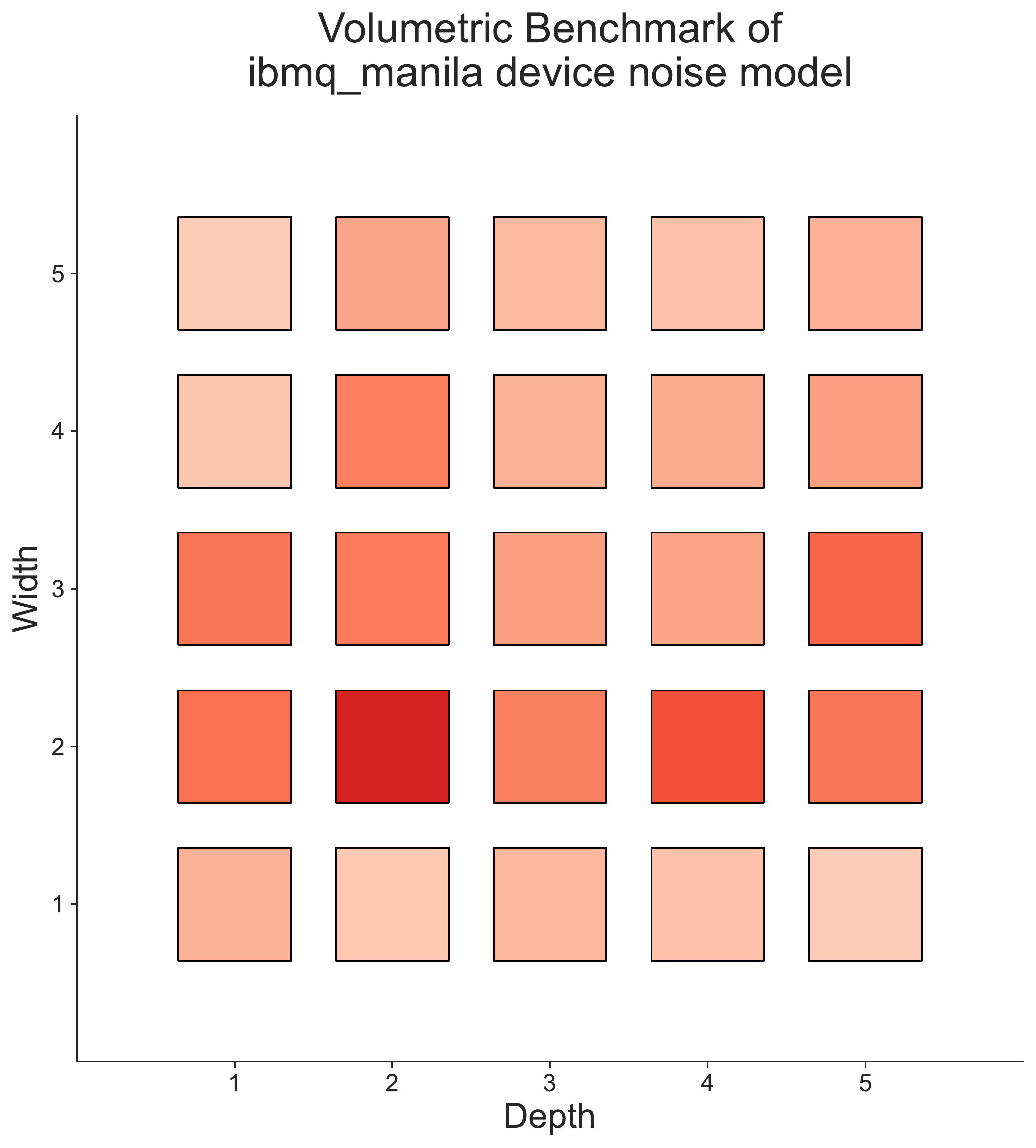}
         \caption{\texttt{ibmq\_manila} device noise model.}
         \label{fig:result_backend}
     \end{subfigure}
     \begin{subfigure}[b]{0.45\textwidth}
         \centering
         \includegraphics[height=0.3\textheight]{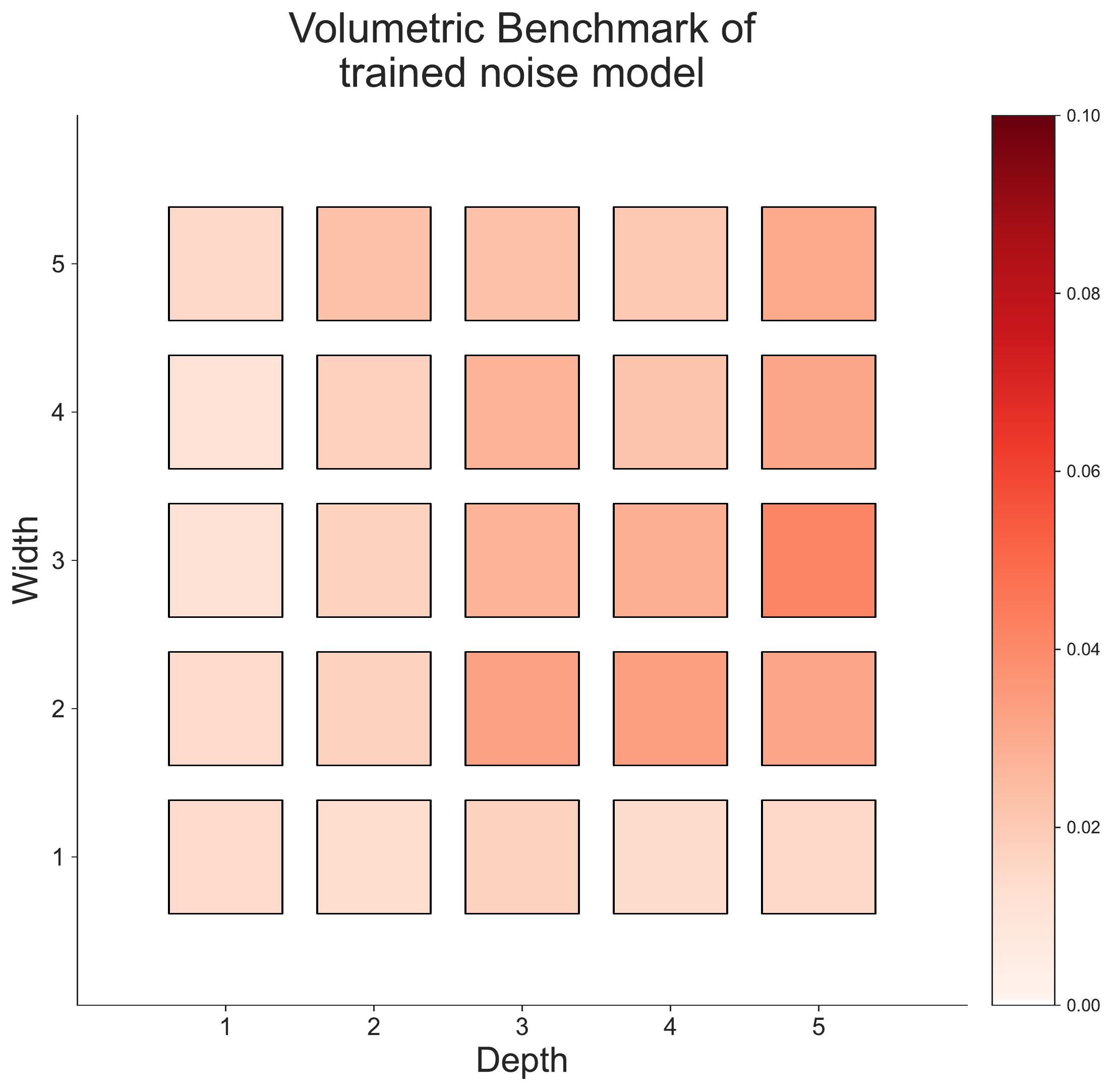}
         \caption{Trained noise model.}
         \label{fig:result_trained}
     \end{subfigure}
     \caption{Volumetric benchmark results for the \texttt{ibmq\_manila} device noise model and the trained noise model.}
     \label{fig:results}
\end{figure*}

Figure~\ref{fig:results} shows the volumetric benchmark results for the device model and our trained model, where the former can be found in figure~\ref{fig:result_backend} and the latter in figure~\ref{fig:result_trained}.\par

\begin{figure*}
\centering
     \begin{subfigure}[b]{0.49\textwidth}
         \centering
         \includegraphics[width=\textwidth]{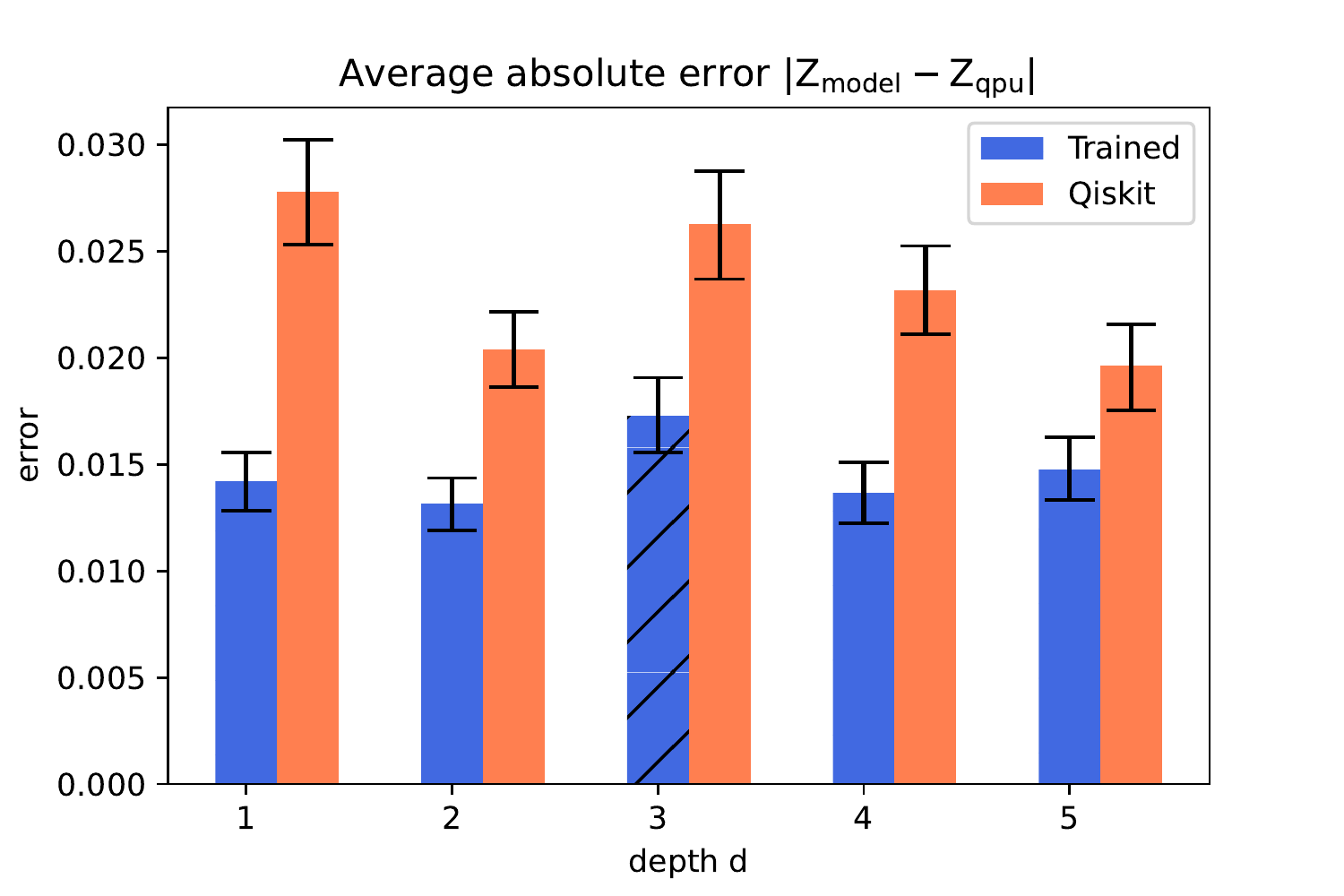}
         \caption{$w=1$}
         \label{fig:w=1}
     \end{subfigure}
     \begin{subfigure}[b]{0.49\textwidth}
         \centering
         \includegraphics[width=\textwidth]{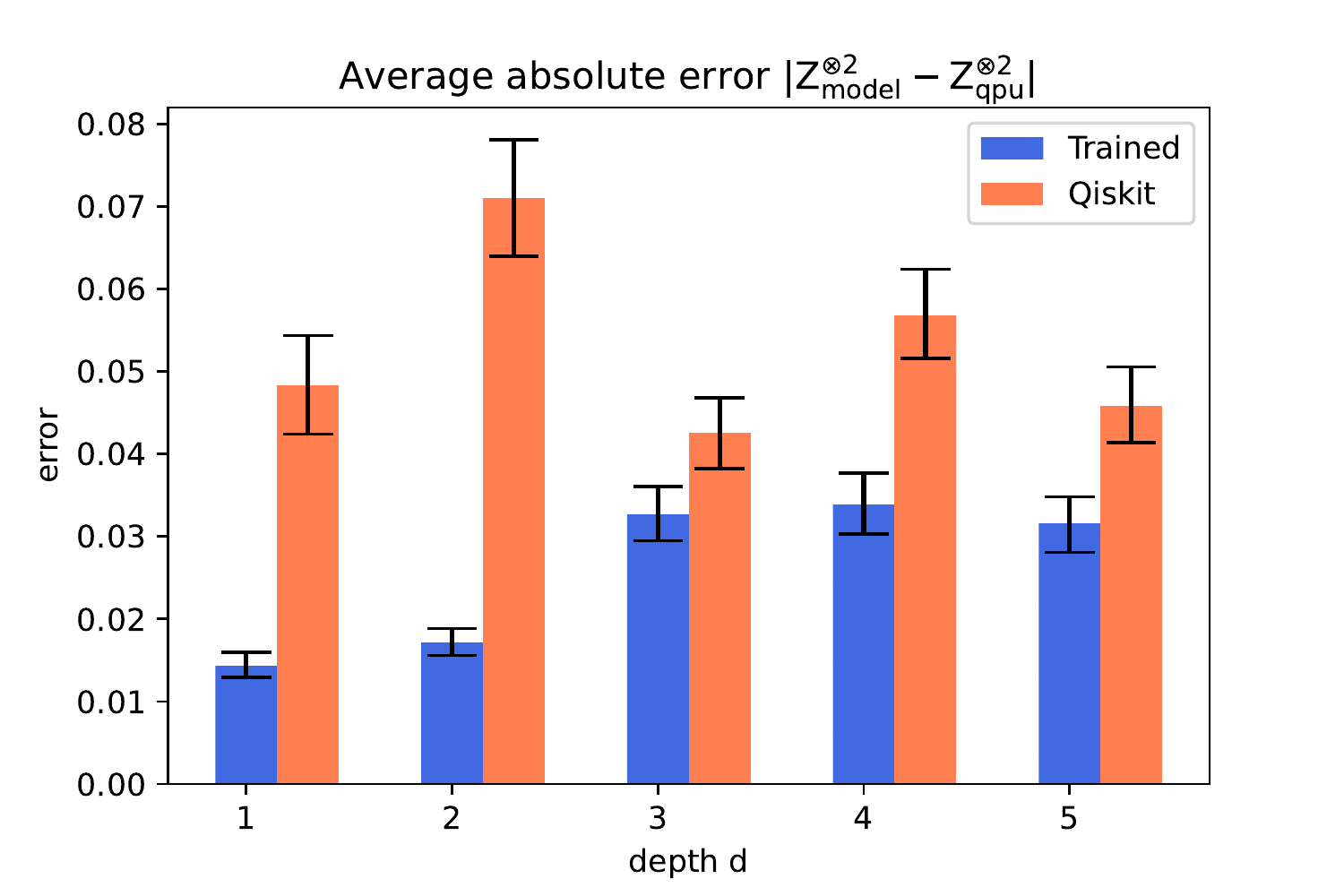}
         \caption{$w=2$}
         \label{fig:w=2}
     \end{subfigure}
     \begin{subfigure}[b]{0.49\textwidth}
         \centering
         \includegraphics[width=\textwidth]{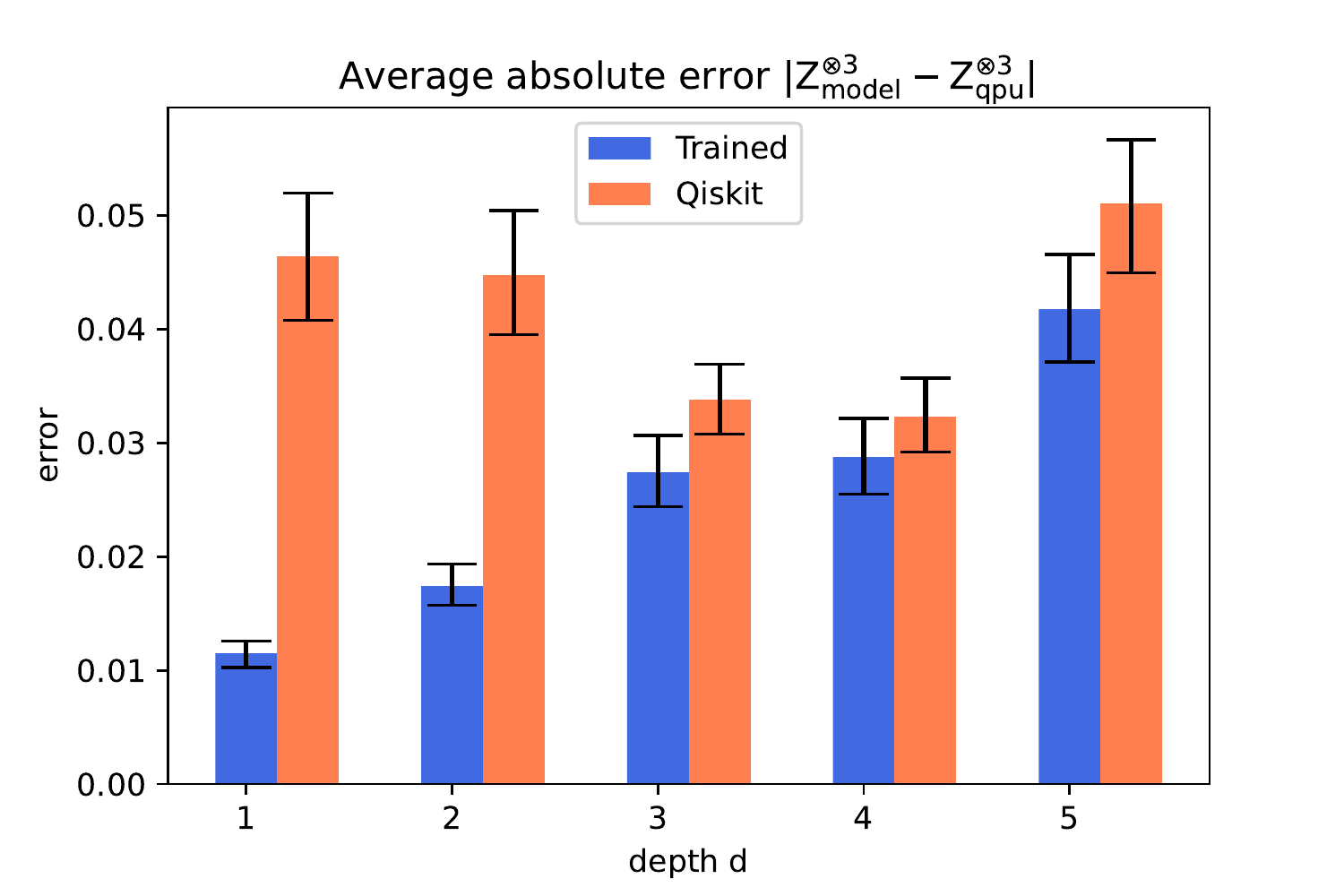}
         \caption{$w=3$}
         \label{fig:w=3}
     \end{subfigure}
     \begin{subfigure}[b]{0.49\textwidth}
         \centering
         \includegraphics[width=\textwidth]{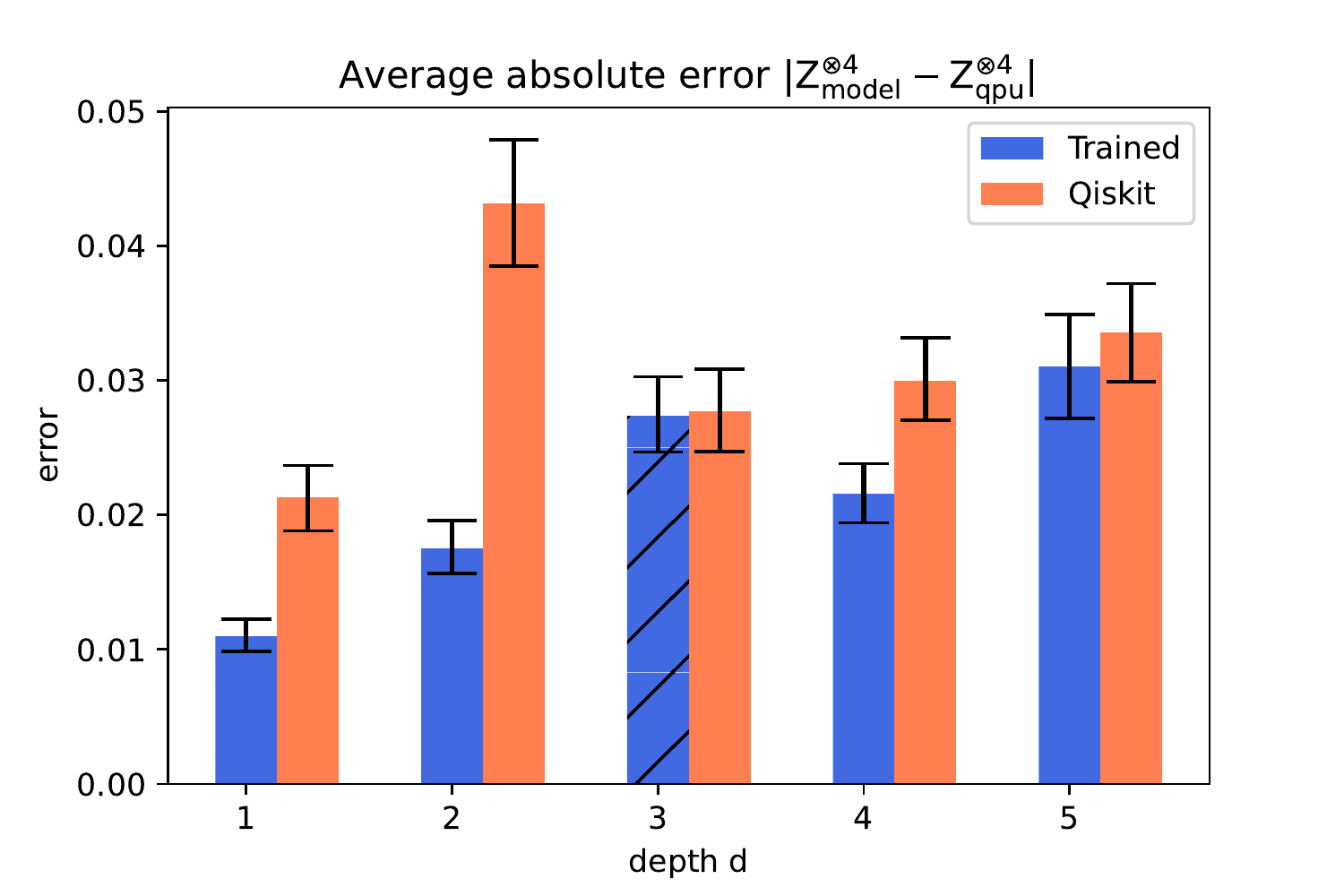}
         \caption{$w=4$}
         \label{fig:w=4}
     \end{subfigure}
     \begin{subfigure}[b]{0.49\textwidth}
         \centering
         \includegraphics[width=\textwidth]{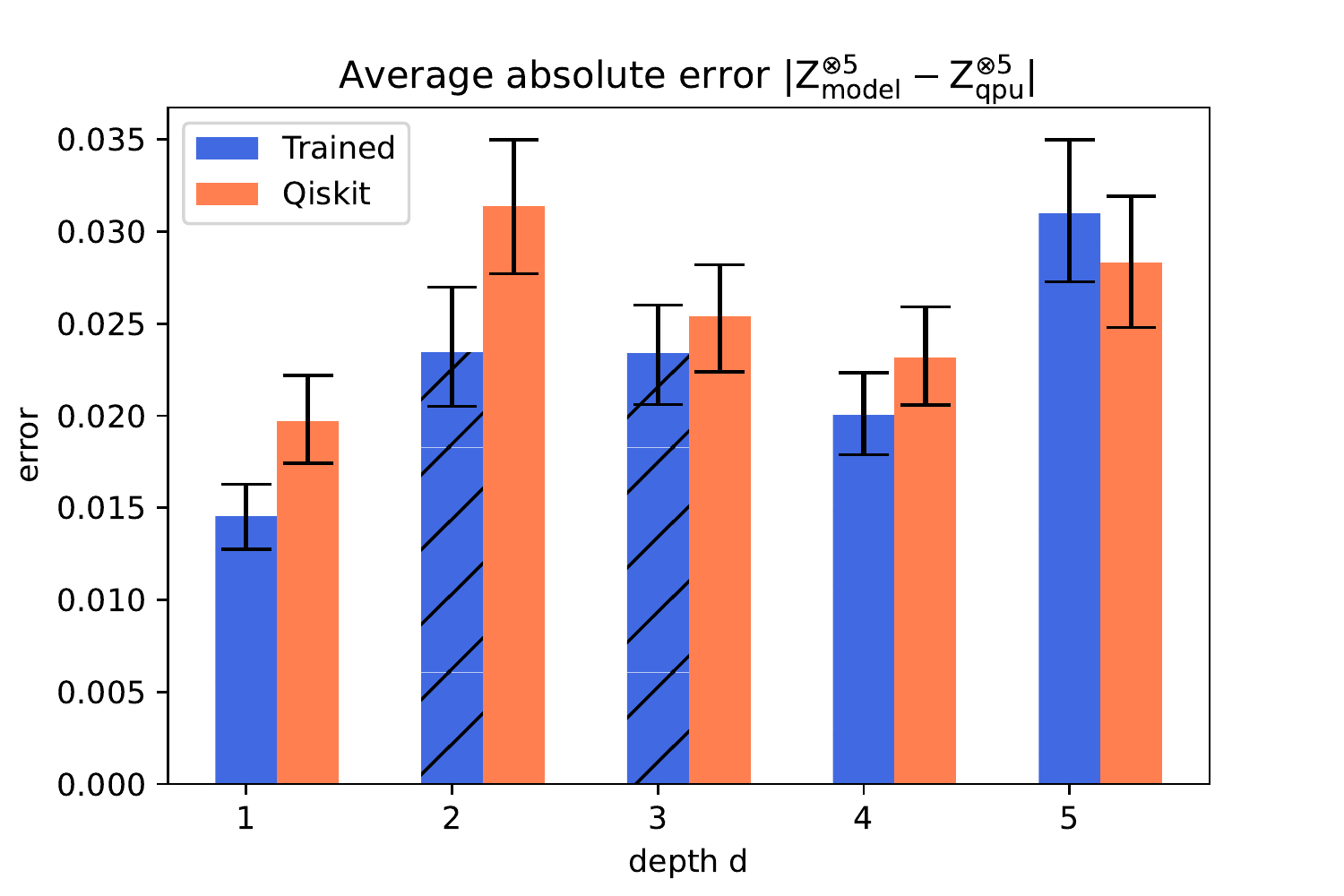}
         \caption{$w=5$}
         \label{fig:w=5}
     \end{subfigure}
     \caption{$95\%$ confidence intervals for average absolute error of $\mathsf{Z}^{\otimes w}$ expectation value. The blue, striped bars represent the results for our trained noise model, while the orange bars show the results for the qiskit device noise model. The x-axes of the plots indicate the depth $d$ of the quantum circuits.}
     \label{fig:confidenceintervals}
\end{figure*}

Figure~\ref{fig:confidenceintervals} shows the confidence intervals of each benchmark based on the procedure from section~\ref{subsec:confidenceintervals}. The blue, striped bars represent our trained noise model, while the red bars show the results of the \texttt{ibmq\_manila} device noise model.

\subsection{Discussion}
\label{subsec:discussion}
The three noise models perform very differently in the volumetric benchmark. Their results improve with model complexity, meaning that our trained model achieves the best results, followed by the qiskit device noise model. In the following, we discuss the volumetric benchmarks in more detail.\par
 
Figure~\ref{fig:result_backend} shows the benchmark results for qiskit’s device noise model. Except for some negative outliers, such as for $w=2$ and $d\in\lbrace2,4\rbrace$, the accuracy of model predictions remains stable for different configurations $(w,d)$. The results strongly depend on the calibration procedure. If parameters are calibrated incorrectly, the deviation between noisy simulation and hardware experiments increases. This is one possible reason for the outliers mentioned above. The device noise model can provide an easily accessible way to simulate quantum circuits with a certain confidence. However, its accuracy is not optimal for realistic simulations. This could change in later versions of qiskit with more error types included.\par

As shown in figure~\ref{fig:result_trained}, our noise model with optimized parameters achieves good overall benchmark results. Its worst performance is an average deviation in $\mathsf{Z}^{\otimes w}$ expectation value of 0.043 (compared to 0.067 of the previous model). The model works particularly well for shallower quantum circuits with up to three layers. For deeper circuits, we observe a slight decrease in its accuracy. This decrease is not necessarily due to a lack of model complexity but could also be caused by suboptimal training.\par

The good performance of our model is also supported by figure~\ref{fig:confidenceintervals}. For all shallow quantum circuits with $d\leq2$, as well as for small qubit numbers with $w\leq 2$, it shows a significant improvement compared to the device noise model. For all other configurations, our model either performs better or equally well within the statistical confidence.

\subsection{Limitations and threats to validity}
\label{subsec:limitations}

In the following, we discuss limitations and threats to validity concerning both contributions from section~\ref{subsec:contribution}.

\subsubsection*{The framework}
The main limitation of our benchmarking framework for noise models is that it does not automatically ensure quality in terms of the criteria from section~\ref{subsec:qualitycriteria}. This quality depends on user choices for test circuits, evaluation metrics, and other specifications.\par
However, we explained in detail in section~\ref{subsec:ensuringquality} how these choices can be made to improve the benchmark quality for each individual criterion.

\subsubsection*{The benchmarks}
There are three threats to validity that we identify for our benchmarks. These threats potentially affect the quality criteria relevance, reproducibility, and fairness.\par

Firstly, hardware results and predictions from the device noise model are obtained using a finite number of shots. Thus, the benchmark results are subject to statistical noise and repeating the benchmarks can yield different outcomes. We mitigate this threat with a large number of shots and quantum circuits.\par

Secondly, the hardware experiments were conducted at different times. Since the noise level in a quantum computer is not constant, the ideal noise model is not always the same. Therefore, comparing the benchmark results of a noise model for one configuration $(w,d)$ to another is not necessarily meaningful. Instead, one should compare the results of different noise models for fixed $(w,d)$.\par

Thirdly, the quantum circuits used for the benchmarks are specific to variational algorithms. Our results are not necessarily generalizable to other applications of quantum computing that use different types of circuits, for example for factorization or search algorithms.

\section{Conclusion}
\label{sec:conclusions}

\subsection{Summary}
\label{subsec:summary}
This paper presents a novel approach to evaluate the accuracy of quantum computing noise models. The approach is based on volumetric benchmarks that compare model predictions to the behaviour of a quantum device for sets of quantum circuits of different sizes. If a noise model performs well in these volumetric benchmarks, it can be used for noisy simulations, reducing the need for quantum hardware. Possibilities to improve the benchmark quality in terms of established quality criteria are also discussed.\par

We conducted volumetric benchmarks for three noise models using the \texttt{ibmq\_manila} quantum computer. The first noise model only considers readout error with calibrated probabilities. The second is the device noise model for the \texttt{ibmq\_manila} hardware from the qiskit library. We construct a third model with trainable parameters that we optimize using a set of training circuits. It contains SPAM error, depolarizing error, thermal relaxation and dephasing, and a simple form of crosstalk error. More types of noise can easily be added to the model.\par

While the readout noise model performed poorly for more than a single qubit, the device and the trained noise model achieved better results for larger system sizes. The predictions of the former still showed larger deviations from hardware data for several configurations of width $w$ and depth $d$. In particular, the accuracy for the configurations $w=2,3$ is decreased. The trained noise model performs significantly better for small qubit numbers ($w\leq2$). Except for the configurations $(3,4)$, $(3,5)$, $(4,3)$, $(4,5)$, and for $w=5, d\geq 3$, where no statistically significant statement can be made, it shows improved results compared to the device noise model. Overall, its accuracy is stable for most configurations. Only for deep quantum circuits do we find a slight decrease. The reason could be a more demanding training environment. Overall, our noise model and approach to training its parameters show promising results in these first volumetric benchmarks.

\subsection{Future work}
\label{subsec:futurework}
The noise model constructed in this paper includes a simple form of crosstalk error. As explained in section~\ref{subsec:quantumnoise} and in more detail in~\cite{sarovar_detecting_2020}, crosstalk can be very versatile. Therefore, future research should construct noise models with more complex variants to further improve our understanding of quantum noise. Additionally, other types of noise could be considered.\par

Furthermore, future research should conduct more extensive volumetric benchmarks. This includes quantum hardware with more qubits and quantum circuits from a larger variety of applications.\par

Moreover, new evaluation criteria for volumetric benchmarks should be investigated to explore other quantum computing applications. While the expectation value of observables is of interest for VQE, different variables are more significant for other algorithms such as Grover~\cite{grover_fast_1996}.\par

Finally, the training method presented in this paper can be improved for better parameter optimization of noise models.

\appendix

\section{Density operators}
\label{sef:densityoperators}

A quantum system can be described by a Hilbert space $\mathcal{H}$ and a bounded, self-adjoint operator $\rho\colon\mathcal{H}\to\mathcal{H}$ called the \textit{density operator} (or \textit{density matrix}). It is defined to satisfy the following properties:
\begin{itemize}
    \item $\rho$ is positive.
    \item $\rho$ is trace class with $\mathsf{tr}(\rho)=1$.
\end{itemize}

The evolution of such a system with a unitary operator $U$ can be expressed as a mapping
\begin{align*}
    \rho\mapsto U\rho U^\dagger\ .
\end{align*}
Measurements are described by a set $\lbrace M_i\rbrace_{i\in\mathcal{I}}$ of measurement operators such that
\begin{align*}
    \sum\limits_{i\in\mathcal{I}}M_i^\dagger M_i = \mathsf{I}\ .
\end{align*}
For any observable $A\colon\mathcal{H}\to\mathcal{H}$, its quantum mechanical \textit{expectation value} is given by

\begin{align*}
     \expval{A}_{\rho}=\mathsf{tr}(A\rho)\ .
\end{align*}
Moreover, if the system is in state $\rho_i$ with probability $p_i$, then its density operator is

\begin{align*}
    \rho = \sum\limits_i p_i\rho_i\ .
\end{align*}
For quantum computing, density operators can be applied as follows. As qubits are two-dimensional quantum systems, they are described by a two-dimensional Hilbert space $\mathcal{H}\simeq\mathbb{C}^2$ with the so-called \textit{computational basis} $\lbrace\ket{0},\ket{1}\rbrace$, where
\begin{align*}
   \ket{0}=\begin{pmatrix}
   1\\0
   \end{pmatrix},\quad
   \ket{1}=\begin{pmatrix}
   0\\1
   \end{pmatrix}\ .
\end{align*}
A qubit state can then be expressed in terms of a $2\times2$ density matrix $\rho$ with $\mathsf{tr}(\rho) = 1$. In quantum computing, qubits are prepared in an initial state, manipulated by unitary gates, and finally measured in the computational basis. Typically, the initial state is $\ket{0}$. The corresponding density matrix is
\begin{align*}
    \rho_0 = \dyad{0}{0}=\begin{pmatrix}
    1&0\\
    0&0
    \end{pmatrix}\ .
\end{align*}
For a composite system of $N$ qubits, the Hilbert space and initial state are $\mathcal{H}=\mathbb{C}^{2N}$ and $\rho_0=\dyad{0\cdots0}$, respectively.

\section{Quantum operations}
\label{sec:quantumoperations}

The term \textit{quantum noise} labels all processes not part of the intended quantum circuit consisting of state preparation, gate operations, and measurements. Quantum operations are a powerful tool for expressing these processes in terms of density operators. Roughly speaking, a quantum operation $\mathcal{E}$ maps the density operator $\rho$ of a quantum system with Hilbert space $\mathcal{H}$ to a density operator $\rho^\prime$ of $\mathcal{H}^\prime$: $\rho^\prime = \mathcal{E}(\rho)$.\par

Mathematically, a quantum operation $\mathcal{E}$ from a  Hilbert space $\mathcal{H}$ to a Hilbert space $\mathcal{H}^\prime$ is a linear map  between their sets of positive trace class operators such that
\begin{itemize}
    \item if $\rho$ is a density operator, then $\mathsf{tr}(\mathcal{E}(\rho))\leq 1$
    \item $\mathcal{E}$ is completely positive.
\end{itemize}

We do not discuss this definition in more detail, instead we refer to the literature for further reading~\cite{nielsen_quantum_2010}.\par

Kraus' theorem~\cite{choi_completely_1975} gives a helpful characterization of quantum operations. It states that a linear map $\mathcal{E}$ between the spaces mentioned above is a quantum operation if and only if there is a set of linear operators $\lbrace O_i\colon \mathcal{H}\to\mathcal{H}^\prime\rbrace$ such that
\begin{align*}
    \mathcal{E}(\rho) = \sum\limits_i O_i\rho O_i^\dagger
\end{align*}
with $\sum_i O_i^\dagger O_i\leq\mathsf{I}$. An important example of a quantum operation for quantum computing is the \textit{depolarizing error}
\begin{align*}
    \mathcal{D}(\rho) = (1-\lambda)\rho + \frac{\lambda}{D}\cdot\mathsf{I}\ ,
\end{align*}
where $D$ is the dimension of the system, i.e. $D=2^N$ for $N$ qubits. Denoting the Pauli matrices by $\mathsf{X}$, $\mathsf{Y}$ and $\mathsf{Z}$, the depolarizing error on a single qubits takes the following form in terms of Kraus operators:
\begin{align*}
    \mathcal{D}(\rho) = \left(1-\frac{3}{4}\lambda\right)\rho + \frac{\lambda}{4}\left(\mathsf{X}\rho\mathsf{X}+\mathsf{Y}\rho\mathsf{Y}+\mathsf{Z}\rho\mathsf{Z}\right)\ .
\end{align*}
\begin{table*}[t]
    \centering
    \begin{tabular}{|c|c|c|c|c|c|}
    \hline
    &$d=1$&$d=2$&$d=3$&$d=4$&$d=5$\\
    \hline
    $w=1$&26th, 19:58&26th, 22:23&27th, 01:08&27th, 02:44&27th, 05:36\\
    \hline
    $w=2$&27th, 10:36&27th, 14:47&27th, 17:56&27th, 21:46&28th, 00:17\\
    \hline
     $w=3$&28th, 03:50&28th, 08:31&28th, 10:35&28th, 16:52&28th, 20:28\\
    \hline
     $w=4$&29th, 12:29&29th, 15:16&29th, 17:39&29th, 20:38&29th, 23:52\\
    \hline
     $w=5$&30th, 02:33&30th, 04:45&30th, 07:56&30th, 10:33&30th, 12:44\\
    \hline
    \end{tabular}
    \caption{Execution times of hardware experiments for volumetric benchmark. All experiments were run on the \texttt{ibmq\_manila device} in September 2022. The times are in UTC+2.}
    \label{tab:running_times}
\end{table*}
\section{Implementation details}
\label{sec:implementation}
This section explains the implementation of the training and benchmarking of noise models in more detail. It contains three parts that discuss simulating quantum circuits, running experiments on quantum hardware, and optimizing parameters.

\subsection{Hardware experiments}
For 25 possible configurations of $(w,d)$, 100 training circuits and 200 benchmark circuits were run on the \texttt{ibmq\_manila} quantum computer. The circuits of a pair $(w,d)$ were first compiled into native gates using the \texttt{transpile} method with the argument \texttt{optimization\_level=2} from the qiskit software library (qiskit version 0.38.0). Afterwards, they were sent to the device as one job and executed consecutively. The running times of the experiments can be found in table~\ref{tab:running_times}. Snapshots of the qiskit device noise model are saved at every run.

\subsection{Noisy simulations}
Similar to the hardware experiments, all circuits are compiled into native gates. The compiled circuits are then simulated with different noise models. We use two software packages for the Python programming languages for these simulations.\par

Simulations of the ibmq\_manila device noise model are implemented using qiskit and its \texttt{AerSimulator} device with the noise model saved at the corresponding hardware run. We always use 8192 shots.\par

Simulations of our noise model are implemented using Pennylane's \texttt{default.mixed} device. This device allows for exact computations of the density matrix and, therefore, for exact predictions of outcome probabilities or expectation values. During the training of the noise model, the outcome distribution is computed using \texttt{probs} measurements. The $\mathsf{Z}^{\otimes w}$ expectation value is calculated with the \texttt{expval} measurement for the volumetric benchmarks.

\subsection{Parameter training}
The noise model parameters are trained using the SPSA algorithm. At every iteration of the optimization process, the loss function from section~\ref{subsec:training} is evaluated on 100 \texttt{EfficientSU2} quantum circuits with randomized parameters. The Hellinger distance is computed by comparing the probability distribution of the noisy simulation to the counts of the hardware run. The latter are interpreted as distribution via their relative frequencies.\par

The optimizer trains for 500 epochs with the hyperparameter $c$ set to $c=0.005$. The $a$ hyperparameter varies between $a=0.005$ and $a=0.08$, depending on $w$ and $d$. Moreover, we use $\alpha=0.602$ and $\gamma=0.101$, as recommended in~\cite{spall_implementation_1998}.

\section*{Acknowledgements}
We acknowledge the support by DASHH (Data Science in Hamburg - HELMHOLTZ Graduate School for the Structure of Matter) with the Grant-No. HIDSS-0002. This research was funded by Deutsches Elektronen-Synchrotron DESY, a member of the Helmholtz Association (HGF). This work is supported with funds from the Ministry of Science, Research and Culture of the State of Brandenburg within the Centre for Quantum Technologies and Applications (CQTA).

\begin{figure}[H]
    \centering
    \includegraphics[width=0.12\textwidth]{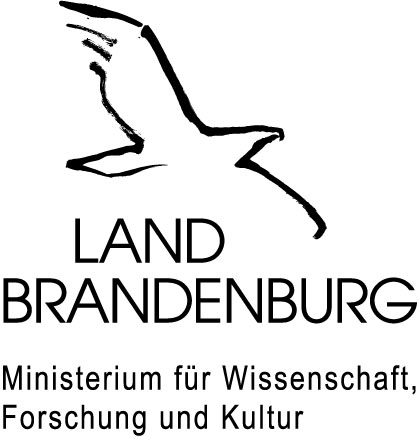}
    \label{fig:enter-label}
\end{figure}

\printbibliography

@article{Sun2021,
	title = {Mitigating Realistic Noise in Practical Noisy Intermediate-Scale Quantum Devices},
	volume = {15},
	url = {https://link.aps.org/doi/10.1103/PhysRevApplied.15.034026},
	doi = {10.1103/PhysRevApplied.15.034026},
	pages = {34026},
	number = {3},
	journaltitle = {Physical Review Applied},
	author = {Sun, Jinzhao and Yuan, Xiao and Tsunoda, Takahiro and Vedral, Vlatko and Benjamin, Simon C and Endo, Suguru},
	date = {2021-03-09},
	note = {Publisher: American Physical Society},
}

@article{vovrosh_simple_2021,
	title = {Simple mitigation of global depolarizing errors in quantum simulations},
	volume = {104},
	url = {https://link.aps.org/doi/10.1103/PhysRevE.104.035309},
	doi = {10.1103/PhysRevE.104.035309},
	number = {3},
	urldate = {2023-04-07},
	journal = {Physical Review E},
	author = {Vovrosh, Joseph and Khosla, Kiran E. and Greenaway, Sean and Self, Christopher and Kim, M. S. and Knolle, Johannes},
	month = sep,
	year = {2021},
	note = {Publisher: American Physical Society},
	pages = {035309},
}

@article{Harper2020,
	title = {Efficient learning of quantum noise},
	volume = {16},
	issn = {17452481},
	doi = {10.1038/s41567-020-0992-8},
	pages = {1184--1188},
	number = {12},
	journaltitle = {Nature Physics},
	author = {Harper, Robin and Flammia, Steven T. and Wallman, Joel J.},
	date = {2020},
	eprinttype = {arxiv},
	eprint = {1907.13022},
}

@article{su_error_2021,
	title = {Error mitigation on a near-term quantum photonic device},
	volume = {5},
	url = {https://quantum-journal.org/papers/q-2021-05-04-452/},
	doi = {10.22331/q-2021-05-04-452},
	language = {},
	urldate = {2023-04-07},
	journal = {Quantum},
	author = {Su, Daiqin and Israel, Robert and Sharma, Kunal and Qi, Haoyu and Dhand, Ish and Brádler, Kamil},
	month = may,
	year = {2021},
	note = {Publisher: Verein zur Förderung des Open Access Publizierens in den Quantenwissenschaften},
	pages = {452},
}

@article{Temme2017,
	title = {Error Mitigation for Short-Depth Quantum Circuits},
	volume = {119},
	issn = {10797114},
	doi = {10.1103/PhysRevLett.119.180509},
	pages = {1--15},
	number = {18},
	journaltitle = {Physical Review Letters},
	author = {Temme, Kristan and Bravyi, Sergey and Gambetta, Jay M.},
	date = {2017},
	pmid = {29219599},
	eprinttype = {arxiv},
	eprint = {1612.02058},
}

@article{Kandala2019,
	title = {Error mitigation extends the computational reach of a noisy quantum processor},
	volume = {567},
	issn = {14764687},
	url = {http://dx.doi.org/10.1038/s41586-019-1040-7},
	doi = {10.1038/s41586-019-1040-7},
	pages = {491--495},
	number = {7749},
	journaltitle = {Nature},
	author = {Kandala, Abhinav and Temme, Kristan and Córcoles, Antonio D. and Mezzacapo, Antonio and Chow, Jerry M. and Gambetta, Jay M.},
	date = {2019},
	pmid = {30918370},
	note = {Publisher: Springer {US}},
}

@article{funcke_measurement_2022,
	title = {Measurement error mitigation in quantum computers through classical bit-flip correction},
	volume = {105},
	url = {https://link.aps.org/doi/10.1103/PhysRevA.105.062404},
	doi = {10.1103/PhysRevA.105.062404},
	pages = {062404},
    year = {2022},
	number = {6},
	journaltitle = {Physical Review A},
	shortjournal = {Phys. Rev. A},
	author = {Funcke, Lena and Hartung, Tobias and Jansen, Karl and Kühn, Stefan and Stornati, Paolo and Wang, Xiaoyang},
	urldate = {2023-04-07},
	date = {2022-06-03},
	note = {Publisher: American Physical Society},
}

@article{Murali2020,
	title = {Software mitigation of crosstalk on noisy intermediate-scale quantum computers},
	doi = {10.1145/3373376.3378477},
	pages = {1001--1016},
	journaltitle = {International Conference on Architectural Support for Programming Languages and Operating Systems - {ASPLOS}},
	author = {Murali, Prakash and {McKay}, David C. and Martonosi, Margaret and Javadi-Abhari, Ali},
	date = {2020},
	eprinttype = {arxiv},
	eprint = {2001.02826},
	note = {{ISBN}: 9781450371025},
}

@book{nielsen_quantum_2010,
	address = {Cambridge ; New York},
	edition = {10th anniversary ed},
	title = {Quantum computation and quantum information},
	isbn = {978-1-107-00217-3},
	publisher = {Cambridge University Press},
	author = {Nielsen, Michael A. and Chuang, Isaac L.},
	year = {2010},
}

@article{choi_completely_1975,
	title = {Completely positive linear maps on complex matrices},
	volume = {10},
	issn = {0024-3795},
	url = {https://www.sciencedirect.com/science/article/pii/0024379575900750},
	doi = {10.1016/0024-3795(75)90075-0},
	pages = {285--290},
	number = {3},
	journaltitle = {Linear Algebra and its Applications},
	shortjournal = {Linear Algebra and its Applications},
	author = {Choi, Man-Duen},
	urldate = {2021-11-18},
	date = {1975-06-01},
	langid = {english},
}

@article{georgopoulos_modelling_2021,
	title = {Modelling and Simulating the Noisy Behaviour of Near-term Quantum Computers},
	volume = {104},
	issn = {2469-9926, 2469-9934},
	url = {http://arxiv.org/abs/2101.02109},
	doi = {10.1103/PhysRevA.104.062432},
	pages = {062432},
	number = {6},
	journaltitle = {Physical Review A},
	shortjournal = {Phys. Rev. A},
	author = {Georgopoulos, Konstantinos and Emary, Clive and Zuliani, Paolo},
	urldate = {2022-01-05},
	date = {2021-12-17},
	eprinttype = {arxiv},
	eprint = {2101.02109},
}

@article{blume-kohout_volumetric_2020,
	title = {A volumetric framework for quantum computer benchmarks},
	volume = {4},
	url = {https://quantum-journal.org/papers/q-2020-11-15-362/},
	doi = {10.22331/q-2020-11-15-362},
	pages = {362},
	journaltitle = {Quantum},
	author = {Blume-Kohout, Robin and Young, Kevin C.},
	urldate = {2022-01-05},
	date = {2020-11-15},
	langid = {british},
	note = {Publisher: Verein zur Förderung des Open Access Publizierens in den Quantenwissenschaften},
}

@article{blume-kohout_wildcard_2020,
	title = {Wildcard error: Quantifying unmodeled errors in quantum processors},
	url = {http://arxiv.org/abs/2012.12231},
	shorttitle = {Wildcard error},
	journaltitle = {{arXiv}:2012.12231 },
	author = {Blume-Kohout, Robin and Rudinger, Kenneth and Nielsen, Erik and Proctor, Timothy and Young, Kevin},
	urldate = {2022-01-12},
	date = {2020-12-22},
	eprinttype = {arxiv},
	eprint = {2012.12231},
}

@article{knill_randomized_2008,
	title = {Randomized benchmarking of quantum gates},
	volume = {77},
	url = {https://link.aps.org/doi/10.1103/PhysRevA.77.012307},
	doi = {10.1103/PhysRevA.77.012307},
	pages = {012307},
	number = {1},
	journaltitle = {Physical Review A},
	shortjournal = {Phys. Rev. A},
	author = {Knill, E. and Leibfried, D. and Reichle, R. and Britton, J. and Blakestad, R. B. and Jost, J. D. and Langer, C. and Ozeri, R. and Seidelin, S. and Wineland, D. J.},
	urldate = {2022-01-12},
	date = {2008-01-08},
	note = {Publisher: American Physical Society},
}

@article{sarovar_detecting_2020,
	title = {Detecting crosstalk errors in quantum information processors},
	volume = {4},
	url = {https://quantum-journal.org/papers/q-2020-09-11-321/},
	doi = {10.22331/q-2020-09-11-321},
	pages = {321},
	journaltitle = {Quantum},
	author = {Sarovar, Mohan and Proctor, Timothy and Rudinger, Kenneth and Young, Kevin and Nielsen, Erik and Blume-Kohout, Robin},
	urldate = {2022-01-17},
	date = {2020-09-11},
	langid = {british},
	note = {Publisher: Verein zur Förderung des Open Access Publizierens in den Quantenwissenschaften},
}

@article{proctor_measuring_2022,
	title = {Measuring the capabilities of quantum computers},
	volume = {18},
	rights = {2021 The Author(s), under exclusive licence to Springer Nature Limited},
	issn = {1745-2481},
	url = {https://www.nature.com/articles/s41567-021-01409-7},
	doi = {10.1038/s41567-021-01409-7},
	pages = {75--79},
	number = {1},
	journaltitle = {Nature Physics},
	shortjournal = {Nat. Phys.},
	author = {Proctor, Timothy and Rudinger, Kenneth and Young, Kevin and Nielsen, Erik and Blume-Kohout, Robin},
	urldate = {2022-02-10},
	date = {2022-01},
	langid = {english},
	note = {Number: 1
Publisher: Nature Publishing Group},
}

@inproceedings{v_kistowski_how_2015,
	location = {New York, {NY}, {USA}},
	title = {How to Build a Benchmark},
	isbn = {978-1-4503-3248-4},
	url = {https://doi.org/10.1145/2668930.2688819},
	doi = {10.1145/2668930.2688819},
	series = {{ICPE} '15},
	pages = {333--336},
	booktitle = {Proceedings of the 6th {ACM}/{SPEC} International Conference on Performance Engineering},
	publisher = {Association for Computing Machinery},
	author = {v. Kistowski, Jóakim and Arnold, Jeremy A. and Huppler, Karl and Lange, Klaus-Dieter and Henning, John L. and Cao, Paul},
	urldate = {2022-02-10},
	date = {2015-01-31},
}

@article{cirstoiu_volumetric_2022,
	title = {Volumetric Benchmarking of Error Mitigation with Qermit},
	url = {http://arxiv.org/abs/2204.09725},
	journaltitle = {{arXiv}:2204.09725 },
	author = {Cirstoiu, Cristina and Dilkes, Silas and Mills, Daniel and Sivarajah, Seyon and Duncan, Ross},
	urldate = {2022-05-03},
	date = {2022-04-20},
	eprinttype = {arxiv},
	eprint = {2204.09725},
}

@article{cross_validating_2019,
	title = {Validating quantum computers using randomized model circuits},
	volume = {100},
	url = {https://link.aps.org/doi/10.1103/PhysRevA.100.032328},
	doi = {10.1103/PhysRevA.100.032328},
	pages = {032328},
	number = {3},
	journaltitle = {Physical Review A},
	shortjournal = {Phys. Rev. A},
	author = {Cross, Andrew W. and Bishop, Lev S. and Sheldon, Sarah and Nation, Paul D. and Gambetta, Jay M.},
	urldate = {2022-06-30},
	date = {2019-09-20},
	note = {Publisher: American Physical Society},
}

@article{emerson_scalable_2005,
	title = {Scalable noise estimation with random unitary operators},
	volume = {7},
	issn = {1464-4266},
	url = {https://doi.org/10.1088/1464-4266/7/10/021},
	doi = {10.1088/1464-4266/7/10/021},
	pages = {S347--S352},
	number = {10},
	journaltitle = {Journal of Optics B: Quantum and Semiclassical Optics},
	shortjournal = {J. Opt. B: Quantum Semiclass. Opt.},
	author = {Emerson, Joseph and Alicki, Robert and Karol Życzkowski},
	urldate = {2022-07-01},
	date = {2005-09},
	langid = {english},
	note = {Publisher: {IOP} Publishing},
}

@inproceedings{grover_fast_1996,
	location = {New York, {NY}, {USA}},
	title = {A fast quantum mechanical algorithm for database search},
	isbn = {978-0-89791-785-8},
	url = {https://doi.org/10.1145/237814.237866},
	doi = {10.1145/237814.237866},
	series = {{STOC} '96},
	pages = {212--219},
	booktitle = {Proceedings of the twenty-eighth annual {ACM} symposium on Theory of Computing},
	publisher = {Association for Computing Machinery},
	author = {Grover, Lov K.},
	urldate = {2023-02-14},
	date = {1996-07-01},
}

@article{zurek_decoherence_1991,
	title = {Decoherence and the {Transition} from {Quantum} to {Classical}},
	volume = {44},
	issn = {0031-9228},
	url = {https://physicstoday.scitation.org/doi/10.1063/1.881293},
	doi = {10.1063/1.881293},
	number = {10},
	urldate = {2022-11-03},
	journal = {Physics Today},
	author = {Zurek, Wojciech H.},
	month = oct,
	year = {1991},
	note = {Publisher: American Institute of Physics},
	pages = {36--44},
}

@article{Peruzzo2014,
	title = {A variational eigenvalue solver on a photonic quantum processor},
	volume = {5},
	issn = {20411723},
	doi = {10.1038/ncomms5213},
	number = {2},
	journal = {Nature Communications},
	author = {Peruzzo, Alberto and McClean, Jarrod and Shadbolt, Peter and Yung, Man Hong and Zhou, Xiao Qi and Love, Peter J. and Aspuru-Guzik, Alán and O'Brien, Jeremy L.},
	year = {2014},
	note = {arXiv: 1304.3061v1},
	pages = {1--10},
}

@misc{ Qiskit,
       author = {MD SAJID ANIS and Abby-Mitchell and H{\'e}ctor Abraham and AduOffei and Rochisha Agarwal and Gabriele Agliardi and Merav Aharoni and Vishnu Ajith and Ismail Yunus Akhalwaya and Gadi Aleksandrowicz and Thomas Alexander and Matthew Amy and Sashwat Anagolum and Anthony-Gandon and Israel F. Araujo and Eli Arbel and Abraham Asfaw and Anish Athalye and Artur Avkhadiev and Carlos Azaustre and PRATHAMESH BHOLE and Vishal Bajpe and Abhik Banerjee and Santanu Banerjee and Will Bang and Aman Bansal and Panagiotis Barkoutsos and Ashish Barnawal and George Barron and George S. Barron and Luciano Bello and Yael Ben-Haim and M. Chandler Bennett and Daniel Bevenius and Dhruv Bhatnagar and Prakhar Bhatnagar and Arjun Bhobe and Paolo Bianchini and Lev S. Bishop and Carsten Blank and Sorin Bolos and Soham Bopardikar and Samuel Bosch and Sebastian Brandhofer and Brandon and Sergey Bravyi and Nick Bronn and Bryce-Fuller and David Bucher and Artemiy Burov and Fran Cabrera and Padraic Calpin and Lauren Capelluto and Jorge Carballo and Gin{\'e}s Carrascal and Adam Carriker and Ivan Carvalho},
       title = {Qiskit: An Open-source Framework for Quantum Computing},
       year = {2021},
       doi = {10.5281/zenodo.2573505},
}

@misc{https://doi.org/10.48550/arxiv.1811.04968,
  doi = {10.48550/ARXIV.1811.04968},
  
  url = {https://arxiv.org/abs/1811.04968},
  
  author = {Bergholm, Ville and Izaac, Josh and Schuld, Maria and Gogolin, Christian and Ahmed, Shahnawaz and Ajith, Vishnu and Alam, M. Sohaib and Alonso-Linaje, Guillermo and AkashNarayanan, B. and Asadi, Ali and Arrazola, Juan Miguel and Azad, Utkarsh and Banning, Sam and Blank, Carsten and Bromley, Thomas R and Cordier, Benjamin A. and Ceroni, Jack and Delgado, Alain and Di Matteo, Olivia and Dusko, Amintor and Garg, Tanya and Guala, Diego and Hayes, Anthony and Hill, Ryan and Ijaz, Aroosa and Isacsson, Theodor and Ittah, David and Jahangiri, Soran and Jain, Prateek and Jiang, Edward and Khandelwal, Ankit and Kottmann, Korbinian and Lang, Robert A. and Lee, Christina and Loke, Thomas and Lowe, Angus and McKiernan, Keri and Meyer, Johannes Jakob and Montañez-Barrera, J. A. and Moyard, Romain and Niu, Zeyue and O'Riordan, Lee James and Oud, Steven and Panigrahi, Ashish and Park, Chae-Yeun and Polatajko, Daniel and Quesada, Nicolás and Roberts, Chase and Sá, Nahum and Schoch, Isidor and Shi, Borun and Shu, Shuli and Sim, Sukin and Singh, Arshpreet and Strandberg, Ingrid and Soni, Jay and Száva, Antal and Thabet, Slimane and Vargas-Hernández, Rodrigo A. and Vincent, Trevor and Vitucci, Nicola and Weber, Maurice and Wierichs, David and Wiersema, Roeland and Willmann, Moritz and Wong, Vincent and Zhang, Shaoming and Killoran, Nathan},
  
  title = {PennyLane: Automatic differentiation of hybrid quantum-classical computations},
  
  publisher = {arXiv},
  
  year = {2018},
  
  copyright = {arXiv.org perpetual, non-exclusive license}
}

@article{Spall1992,
	title = {Multivariate {Stochastic} {Approximation} {Using} a {Simultaneous} {Perturbation} {Gradient} {Approximation}},
	volume = {37},
	issn = {15582523},
	doi = {10.1109/9.119632},
	number = {3},
	journal = {IEEE Transactions on Automatic Control},
	author = {Spall, James C.},
	year = {1992},
	pages = {332--341},
}

@article{spall_implementation_1998,
	title = {Implementation of the simultaneous perturbation algorithm for stochastic optimization},
	volume = {34},
	issn = {1557-9603},
	doi = {10.1109/7.705889},
	number = {3},
	journal = {IEEE Transactions on Aerospace and Electronic Systems},
	author = {Spall, J.C.},
	month = jul,
	year = {1998},
	note = {Conference Name: IEEE Transactions on Aerospace and Electronic Systems},
	pages = {817--823},
}

@article{Kwon2020a,
	title = {A {Hybrid} {Quantum}-{Classical} {Approach} to {Mitigating} {Measurement} {Errors} in {Quantum} {Algorithms}},
	volume = {70},
	issn = {1557-9956},
	doi = {10.1109/TC.2020.3009664},
	number = {9},
	journal = {IEEE Transactions on Computers},
	author = {Kwon, Hyeokjea and Bae, Joonwoo},
	month = sep,
	year = {2021},
	note = {Conference Name: IEEE Transactions on Computers},
	pages = {1401--1411},
}

@article{Bravyi2021,
	title = {Mitigating measurement errors in multiqubit experiments},
	volume = {103},
	url = {https://link.aps.org/doi/10.1103/PhysRevA.103.042605},
	doi = {10.1103/PhysRevA.103.042605},
	number = {4},
	journal = {Physical Review A},
	author = {Bravyi, Sergey and Sheldon, Sarah and Kandala, Abhinav and Mckay, David C and Gambetta, Jay M},
	month = apr,
	year = {2021},
	note = {Publisher: American Physical Society},
	pages = {42605},
}

@article{Endo2018,
	title = {Practical {Quantum} {Error} {Mitigation} for {Near}-{Future} {Applications}},
	volume = {8},
	issn = {21603308},
	doi = {10.1103/PhysRevX.8.031027},
	number = {3},
	journal = {Physical Review X},
	author = {Endo, Suguru and Benjamin, Simon C. and Li, Ying},
	year = {2018},
	note = {arXiv: 1712.09271},
	pages = {1--20},
}

@article{Tannu2019a,
	title = {Mitigating measurement errors in quantum computers by exploiting state-dependent bias},
	issn = {10724451},
	doi = {10.1145/3352460.3358265},
	journal = {Proceedings of the Annual International Symposium on Microarchitecture, MICRO},
	author = {Tannu, Swamit S. and Qureshi, Moinuddin K.},
	year = {2019},
	note = {ISBN: 9781450369381},
	pages = {279--290},
}

@article{Preskill2018,
	title = {Quantum computing in the {NISQ} era and beyond},
	volume = {2},
	issn = {2521327X},
	doi = {10.22331/q-2018-08-06-79},
	number = {July},
	journal = {Quantum},
	author = {Preskill, John},
	year = {2018},
	note = {arXiv: 1801.00862},
	pages = {1--20},
}

@article{erhard_characterizing_2019,
	title = {Characterizing large-scale quantum computers via cycle benchmarking},
	volume = {10},
	copyright = {2019 The Author(s)},
	issn = {2041-1723},
	url = {https://www.nature.com/articles/s41467-019-13068-7},
	doi = {10.1038/s41467-019-13068-7},
	language = {},
	number = {1},
	urldate = {2022-07-26},
	journal = {Nature Communications},
	author = {Erhard, Alexander and Wallman, Joel J. and Postler, Lukas and Meth, Michael and Stricker, Roman and Martinez, Esteban A. and Schindler, Philipp and Monz, Thomas and Emerson, Joseph and Blatt, Rainer},
	month = nov,
	year = {2019},
	note = {Number: 1
Publisher: Nature Publishing Group},
	pages = {5347},
}

@article{geller_toward_2021,
	title = {Toward efficient correction of multiqubit measurement errors: pair correlation method},
	volume = {6},
	issn = {2058-9565},
	shorttitle = {Toward efficient correction of multiqubit measurement errors},
	url = {https://dx.doi.org/10.1088/2058-9565/abd5c9},
	doi = {10.1088/2058-9565/abd5c9},
	language = {},
	number = {2},
	urldate = {2022-10-23},
	journal = {Quantum Science and Technology},
	author = {Geller, Michael R. and Sun, Mingyu},
	month = feb,
	year = {2021},
	note = {Publisher: IOP Publishing},
	pages = {025009},
}

@misc{greenbaum_introduction_2015,
	title = {Introduction to {Quantum} {Gate} {Set} {Tomography}},
	url = {http://arxiv.org/abs/1509.02921},
	doi = {10.48550/arXiv.1509.02921},
	urldate = {2022-08-03},
	publisher = {arXiv},
	author = {Greenbaum, Daniel},
	month = sep,
	year = {2015},
	note = {arXiv:1509.02921 },
}

@article{merkel_self-consistent_2013,
	title = {Self-consistent quantum process tomography},
	volume = {87},
	url = {https://link.aps.org/doi/10.1103/PhysRevA.87.062119},
	doi = {10.1103/PhysRevA.87.062119},
	number = {6},
	urldate = {2022-08-03},
	journal = {Physical Review A},
	author = {Merkel, Seth T. and Gambetta, Jay M. and Smolin, John A. and Poletto, Stefano and Córcoles, Antonio D. and Johnson, Blake R. and Ryan, Colm A. and Steffen, Matthias},
	month = jun,
	year = {2013},
	note = {Publisher: American Physical Society},
	pages = {062119},
}

@article{blume-kohout_robust_2013,
	title = {Robust, self-consistent, closed-form tomography of quantum logic gates on a trapped ion qubit},
	url = {http://arxiv.org/abs/1310.4492},
	doi = {10.48550/arXiv.1310.4492},
	urldate = {2022-08-03},
	publisher = {arXiv},
	author = {Blume-Kohout, Robin and Gamble, John King and Nielsen, Erik and Mizrahi, Jonathan and Sterk, Jonathan D. and Maunz, Peter},
	month = oct,
	year = {2013},
	note = {arXiv:1310.4492 },
}

@article{poyatos_complete_1997,
	title = {Complete {Characterization} of a {Quantum} {Process}: {The} {Two}-{Bit} {Quantum} {Gate}},
	volume = {78},
	shorttitle = {Complete {Characterization} of a {Quantum} {Process}},
	url = {https://link.aps.org/doi/10.1103/PhysRevLett.78.390},
	doi = {10.1103/PhysRevLett.78.390},
	number = {2},
	urldate = {2022-08-03},
	journal = {Physical Review Letters},
	author = {Poyatos, J. F. and Cirac, J. I. and Zoller, P.},
	month = jan,
	year = {1997},
	note = {Publisher: American Physical Society},
	pages = {390--393},
}

@article{chuang_prescription_1997,
	title = {Prescription for experimental determination of the dynamics of a quantum black box},
	volume = {44},
	issn = {0950-0340},
	url = {https://www.tandfonline.com/doi/abs/10.1080/09500349708231894},
	doi = {10.1080/09500349708231894},
	number = {11-12},
	urldate = {2022-08-03},
	journal = {Journal of Modern Optics},
	author = {Chuang, Isaac L. and Nielsen, M. A.},
	month = nov,
	year = {1997},
	note = {Publisher: Taylor \& Francis
\_eprint: https://www.tandfonline.com/doi/pdf/10.1080/09500349708231894},
	pages = {2455--2467},
}

@article{leibfried_experimental_1996,
	title = {Experimental {Determination} of the {Motional} {Quantum} {State} of a {Trapped} {Atom}},
	volume = {77},
	url = {https://link.aps.org/doi/10.1103/PhysRevLett.77.4281},
	doi = {10.1103/PhysRevLett.77.4281},
	number = {21},
	urldate = {2022-08-03},
	journal = {Physical Review Letters},
	author = {Leibfried, D. and Meekhof, D. M. and King, B. E. and Monroe, C. and Itano, W. M. and Wineland, D. J.},
	month = nov,
	year = {1996},
	note = {Publisher: American Physical Society},
	pages = {4281--4285},
}

@article{nielsen_gate_2021,
	title = {Gate {Set} {Tomography}},
	volume = {5},
	url = {https://quantum-journal.org/papers/q-2021-10-05-557/},
	doi = {10.22331/q-2021-10-05-557},
	language = {},
	urldate = {2022-08-03},
	journal = {Quantum},
	author = {Nielsen, Erik and Gamble, John King and Rudinger, Kenneth and Scholten, Travis and Young, Kevin and Blume-Kohout, Robin},
	month = oct,
	year = {2021},
	note = {Publisher: Verein zur Förderung des Open Access Publizierens in den Quantenwissenschaften},
	pages = {557},
}

@article{dariano_quantum_2002,
	title = {Quantum tomography as a tool for the characterization of optical devices},
	volume = {4},
	issn = {1464-4266},
	url = {https://dx.doi.org/10.1088/1464-4266/4/3/366},
	doi = {10.1088/1464-4266/4/3/366},
	language = {},
	number = {3},
	urldate = {2022-11-05},
	journal = {Journal of Optics B: Quantum and Semiclassical Optics},
	author = {D'Ariano, G. Mauro and Laurentis, Martina De and Paris, Matteo G. A. and Porzio, Alberto and Solimeno, Salvatore},
	month = mar,
	year = {2002},
	pages = {S127},
}

@article{altepeter_ancilla-assisted_2003,
	title = {Ancilla-{Assisted} {Quantum} {Process} {Tomography}},
	volume = {90},
	url = {https://link.aps.org/doi/10.1103/PhysRevLett.90.193601},
	doi = {10.1103/PhysRevLett.90.193601},
	number = {19},
	urldate = {2022-11-05},
	journal = {Physical Review Letters},
	author = {Altepeter, J. B. and Branning, D. and Jeffrey, E. and Wei, T. C. and Kwiat, P. G. and Thew, R. T. and O’Brien, J. L. and Nielsen, M. A. and White, A. G.},
	month = may,
	year = {2003},
	note = {Publisher: American Physical Society},
	pages = {193601},
}

@article{mohseni_quantum-process_2008,
	title = {Quantum-process tomography: {Resource} analysis of different strategies},
	volume = {77},
	shorttitle = {Quantum-process tomography},
	url = {https://link.aps.org/doi/10.1103/PhysRevA.77.032322},
	doi = {10.1103/PhysRevA.77.032322},
	number = {3},
	urldate = {2022-11-05},
	journal = {Physical Review A},
	author = {Mohseni, M. and Rezakhani, A. T. and Lidar, D. A.},
	month = mar,
	year = {2008},
	note = {Publisher: American Physical Society},
	pages = {032322},
}

@article{shabani_efficient_2011,
	title = {Efficient {Measurement} of {Quantum} {Dynamics} via {Compressive} {Sensing}},
	volume = {106},
	url = {https://link.aps.org/doi/10.1103/PhysRevLett.106.100401},
	doi = {10.1103/PhysRevLett.106.100401},
	number = {10},
	urldate = {2022-11-05},
	journal = {Physical Review Letters},
	author = {Shabani, A. and Kosut, R. L. and Mohseni, M. and Rabitz, H. and Broome, M. A. and Almeida, M. P. and Fedrizzi, A. and White, A. G.},
	month = mar,
	year = {2011},
	note = {Publisher: American Physical Society},
	pages = {100401},
}

@article{emerson_symmetrized_2007,
	title = {Symmetrized {Characterization} of {Noisy} {Quantum} {Processes}},
	volume = {317},
	url = {https://www.science.org/doi/10.1126/science.1145699},
	doi = {10.1126/science.1145699},
	number = {5846},
	urldate = {2022-11-05},
	journal = {Science},
	author = {Emerson, Joseph and Silva, Marcus and Moussa, Osama and Ryan, Colm and Laforest, Martin and Baugh, Jonathan and Cory, David G. and Laflamme, Raymond},
	month = sep,
	year = {2007},
	note = {Publisher: American Association for the Advancement of Science},
	pages = {1893--1896},
}

@article{kullback_information_1951,
	title = {On {Information} and {Sufficiency}},
	volume = {22},
	issn = {0003-4851, 2168-8990},
	url = {https://projecteuclid.org/journals/annals-of-mathematical-statistics/volume-22/issue-1/On-Information-and-Sufficiency/10.1214/aoms/1177729694.full},
	doi = {10.1214/aoms/1177729694},
	number = {1},
	urldate = {2022-11-05},
	journal = {The Annals of Mathematical Statistics},
	author = {Kullback, S. and Leibler, R. A.},
	month = mar,
	year = {1951},
	note = {Publisher: Institute of Mathematical Statistics},
	pages = {79--86},
}

@article{hellinger_neue_1909,
	title = {Neue {Begründung} der {Theorie} quadratischer {Formen} von unendlichvielen {Veränderlichen}.},
	volume = {1909},
	issn = {1435-5345},
	url = {https://www.degruyter.com/document/doi/10.1515/crll.1909.136.210/html},
	doi = {10.1515/crll.1909.136.210},
	language = {},
	number = {136},
	urldate = {2022-11-05},
	journal = {Journal für die reine und angewandte Mathematik},
	author = {Hellinger, E.},
	month = jul,
	year = {1909},
	note = {Publisher: De Gruyter},
	pages = {210--271},
}

@article{sanders_bounding_2015,
	title = {Bounding quantum gate error rate based on reported average fidelity},
	volume = {18},
	issn = {1367-2630},
	url = {https://dx.doi.org/10.1088/1367-2630/18/1/012002},
	doi = {10.1088/1367-2630/18/1/012002},
	language = {},
	number = {1},
	urldate = {2022-11-05},
	journal = {New Journal of Physics},
	author = {Sanders, Yuval R. and Wallman, Joel J. and Sanders, Barry C.},
	month = dec,
	year = {2015},
	note = {Publisher: IOP Publishing},
	pages = {012002},
}

@article{takagi_fundamental_2022,
	title = {Fundamental limits of quantum error mitigation},
	volume = {8},
	copyright = {2022 The Author(s)},
	issn = {2056-6387},
	url = {https://www.nature.com/articles/s41534-022-00618-z},
	doi = {10.1038/s41534-022-00618-z},
	number = {1},
	urldate = {2022-11-06},
	journal = {npj Quantum Information},
	author = {Takagi, Ryuji and Endo, Suguru and Minagawa, Shintaro and Gu, Mile},
	month = sep,
	year = {2022},
	note = {Number: 1
Publisher: Nature Publishing Group},
	pages = {1--11},
}

@misc{gustiani_virtual_2023,
	title = {The Virtual Quantum Device ({VQD}): A tool for detailed emulation of quantum computers},
	url = {http://arxiv.org/abs/2306.07342},
	doi = {10.48550/arXiv.2306.07342},
	shorttitle = {The Virtual Quantum Device ({VQD})},
	number = {{arXiv}:2306.07342},
	publisher = {{arXiv}},
	author = {Gustiani, Cica and Jones, Tyson and Benjamin, Simon C.},
	urldate = {2023-06-14},
	date = {2023-06-12},
	eprinttype = {arxiv},
	eprint = {2306.07342 [quant-ph]},
}

\end{document}